\documentclass[a4paper,5p,fleqn]{cas-dc}
\usepackage[numbers]{natbib}
\usepackage{multirow}
\usepackage[table]{xcolor}
\definecolor{lightgreen}{RGB}{144, 238, 144}
\definecolor{lightorange}{RGB}{255, 200, 100}
\definecolor{lightred}{RGB}{255, 100, 100}

\def\tsc#1{\csdef{#1}{\textsc{\lowercase{#1}}\xspace}}
\tsc{WGM}
\tsc{QE}
\tsc{EP}
\tsc{PMS}
\tsc{BEC}
\tsc{DE}
\begin{document}
\let\WriteBookmarks\relax
\def\floatpagepagefraction{1}
\def\textpagefraction{.001}
% \shorttitle{Leveraging social media news}
\shorttitle{USF-MAE: Ultrasound Self-Supervised Foundation Model}
\shortauthors{Y. Megahed et~al.}

\title [mode = title]{USF-MAE: Ultrasound Self-Supervised Foundation Model with Masked Autoencoding}                      
% \tnotemark[1,2]

% \tnotetext[1]{This document is the results of the research
%    project funded by the National Science Foundation.}

% \tnotetext[2]{The second title footnote which is a longer text matter
%    to fill through the whole text width and overflow into
%    another line in the footnotes area of the first page.}

\author[1,5]{Youssef Megahed}[type=editor,
                        %auid=000,bioid=1,
                        %prefix=Sir,
                        %role=Researcher,
                        orcid=0009-0004-2595-5468]
                        
\cormark[1]
\ead{youssefmegahed@cmail.carleton.ca}
% \ead[url]{www.jkkrishnan.in}

\credit{Conceptualization of this study, Methodology, Software}

\affiliation[1]{organization={Department of Systems and Computer Engineering, Carleton University}, 
                city={Ottawa},
                state={Ontario},
                country={Canada}}
                
\affiliation[2]{organization={Department of Clinical Science and Translational Medicine, University of Ottawa}, 
                city={Ottawa},
                state={Ontario},
                country={Canada}}

\affiliation[3]{organization={Department of Obstetrics and Gynecology, University of Ottawa}, 
                city={Ottawa},
                state={Ontario},
                country={Canada}}

\affiliation[4]{organization={School of Epidemiology and Public Health, University of Ottawa}, 
                city={Ottawa},
                state={Ontario},
                country={Canada}}

\affiliation[5]{organization={Department of Methodological and Implementation Research, Ottawa Hospital Research Institute}, 
                city={Ottawa},
                state={Ontario},
                country={Canada}}

\affiliation[6]{organization={Department of Acute Care Research, Ottawa Hospital Research Institute}, 
                city={Ottawa},
                state={Ontario},
                country={Canada}}

\author[6]{Robin Ducharme}
\author[2,6]{Aylin Erman}
\author[3,6]{Mark Walker}
\author[1,2,4,5]{Steven Hawken}
\cormark[2]
\author[1]{Adrian D. C. Chan}

\cormark[1]
\cormark[2]
% \fnmark[2]
\ead{adrianchan@cunet.carleton.ca}

% \fnmark[2]
% \ead{wjh@example.org}
% \ead[URL]{https://www.university.org}

\credit{Data curation, Writing - Original draft preparation}

% \cormark[2]
% \fnmark[1,3]
% \ead{t.rafeeq@example.in}
% \ead[URL]{www.campus.in}

\cortext[cor1]{Corresponding author}
\cortext[cor2]{Co-advising author}
% \cortext[cor2]{Principal corresponding author}

\begin{abstract}
Ultrasound imaging is one of the most widely used diagnostic modalities, offering real-time, radiation-free assessment across diverse clinical domains. However, interpretation of ultrasound images remains challenging due to high noise levels, operator dependence, and limited field of view, resulting in substantial inter-observer variability. Current Deep Learning approaches are hindered by the scarcity of large labeled datasets and the domain gap between general and sonographic images, which limits the transferability of models pretrained on non-medical data. To address these challenges, we introduce the \textit{Ultrasound Self-Supervised Foundation Model with Masked Autoencoding} (\textbf{USF-MAE}), the first large-scale self-supervised MAE framework pretrained exclusively on ultrasound data. The model was pre-trained on $\sim$370,000 2D and 3D ultrasound images curated from 46 open-source datasets, collectively termed \textbf{OpenUS-46}, spanning over twenty anatomical regions. This curated dataset has been made publicly available to facilitate further research and reproducibility. Using a Vision Transformer encoder-decoder architecture, USF-MAE reconstructs masked image patches, enabling it to learn rich, modality-specific representations directly from unlabeled data. The pretrained encoder was fine-tuned on three public downstream classification benchmarks: BUS-BRA (breast cancer), MMOTU-2D (ovarian tumors), and GIST514-DB (gastrointestinal stromal tumors). Across all tasks, USF-MAE consistently outperformed conventional CNN and ViT baselines, achieving F1-scores of 81.6\%, 79.6\%, and 82.4\%, respectively. Despite not using labels during pretraining, USF-MAE approached the performance of the supervised foundation model UltraSam on breast cancer classification and surpassed it on the other tasks, demonstrating strong cross-anatomical generalization. These findings establish USF-MAE as a scalable and label-efficient ultrasound foundation model. Its ability to continuously pretrain on future unlabeled public or institutional datasets without requiring manual annotation makes it an adaptable and sustainable framework for ultrasound representation learning, supporting data-efficient clinical and research applications.
\end{abstract}

% \begin{graphicalabstract}
% \includegraphics{figs/cas-grabs.pdf}
% \end{graphicalabstract}

% \begin{highlights}
% \item Research highlights item 1
% \item Research highlights item 2
% \item Research highlights item 3
% \end{highlights}


\maketitle

\section{Introduction}
Ultrasound (US) imaging is valuable in modern medicine due to its ability to provide real-time, radiation-free, and cost-effective visualization of internal anatomy{~\hypersetup{hidelinks}\textcolor{blue}{\cite{b1}}}. It plays a critical role in a variety of clinical scenarios, from obstetric fetal monitoring to cardiac, abdominal, and emergency imaging, and is widely accessible even in resource-limited settings. US is among the most frequently used diagnostic imaging modalities; for example, in England’s NHS, it accounted for the second-highest number of imaging procedures (around 0.72 million in one month, behind only X-rays){~\hypersetup{hidelinks}\textcolor{blue}{\cite{b2}}}. In the United States, annual US utilization grew from 38.6 million exams in 2011 to 59.8 million in 2021, a 55\% increase{~\hypersetup{hidelinks}\textcolor{blue}{\cite{b3}}}, underscoring the modality’s expanding importance in healthcare.

Although US imaging has many benefits, its images can be challenging to interpret and analyze. US images often have a low signal-to-noise ratio (SNR) and exhibit speckle noise artifacts{~\hypersetup{hidelinks}\textcolor{blue}{\cite{b4}}}. They show substantial variability and are highly dependent on the skill, experience, and technique of the operator{~\hypersetup{hidelinks}\textcolor{blue}{\cite{b5}}}. In addition, US imaging is limited by a restricted field of view, and structures may be obscured by acoustic shadows or interference from bone and gas. Because of these challenges, US scans frequently require interpretation by expert sonographers or radiologists. However, even expert interpretation is subject to inter- and intra-observer variability, meaning that different clinicians or even the same clinician at different times may provide divergent assessments of the same image{~\hypersetup{hidelinks}\textcolor{blue}{\cite{b5, b6}}}. Such variability introduces subjectivity into diagnostic decision-making, which can lead to inconsistent measurements, delayed diagnoses, or potential misclassification of lesions and pathologies, particularly in subtle or borderline cases.

In recent years, deep learning methods, especially Convolutional Neural Networks (CNNs) and transformer-based architectures, have been applied to medical image analysis for tasks such as organ or tumor segmentation, lesion detection \& classification, and anatomical landmark localization{~\hypersetup{hidelinks}\textcolor{blue}{\cite{b1,b7,b8,b19,b20,b25,b26,b27}}}. By automatically recognizing patterns in US images, these AI-driven algorithms have shown potential to reduce the need for specialized human expertise or improve their efficiency. However, a fundamental hurdle is the requirement of large, annotated datasets for training supervised deep learning models. Obtaining ground truth labels (e.g., delineating tumor boundaries or classifying pathology) for US is labour-intensive and typically demands expert knowledge; as a result, labeled US data tends to be scarce and narrowly distributed{~\hypersetup{hidelinks}\textcolor{blue}{\cite{b1}}}. The scarcity of diverse annotated data leads to models that may not generalize well. 

A common workaround has been to leverage transfer learning from models pretrained on large-scale general image datasets, such as ImageNet, which is comprised of over 1.2 million labeled images of common items, with 1,000 of classes spanning animals, vehicles, furniture, scenes, etc.{~\hypersetup{hidelinks}\textcolor{blue}{\cite{b9}}}. Unfortunately, the domain shift between these general images and sonographic images is substantial. For example, US images have specific textures, speckle noise, acoustic shadows, and low-contrast boundaries, which general image features (edges, colours, textures in daylight scenes) do not capture well{~\hypersetup{hidelinks}\textcolor{blue}{\cite{b1}}}. Consequently, while ImageNet pretraining often yields faster convergence and improved performance over training from scratch{~\hypersetup{hidelinks}\textcolor{blue}{\cite{b10}}}, its features may not generalize optimally to US tasks, and domain-specific pretraining or adaptation may outperform the generic case{~\hypersetup{hidelinks}\textcolor{blue}{\cite{b11,b12}}}.  Even within US, there is significant variability across different anatomical targets (e.g., breast, thyroid, and fetal scans), imaging devices, and processing techniques{~\hypersetup{hidelinks}\textcolor{blue}{\cite{b1}}}. As a result, traditional CNN-based approaches fine-tuned on limited US datasets often struggle to maintain accuracy across varied US downstream tasks and datasets.

\begin{figure*}
	\centering
	\includegraphics[width=0.98\textwidth]{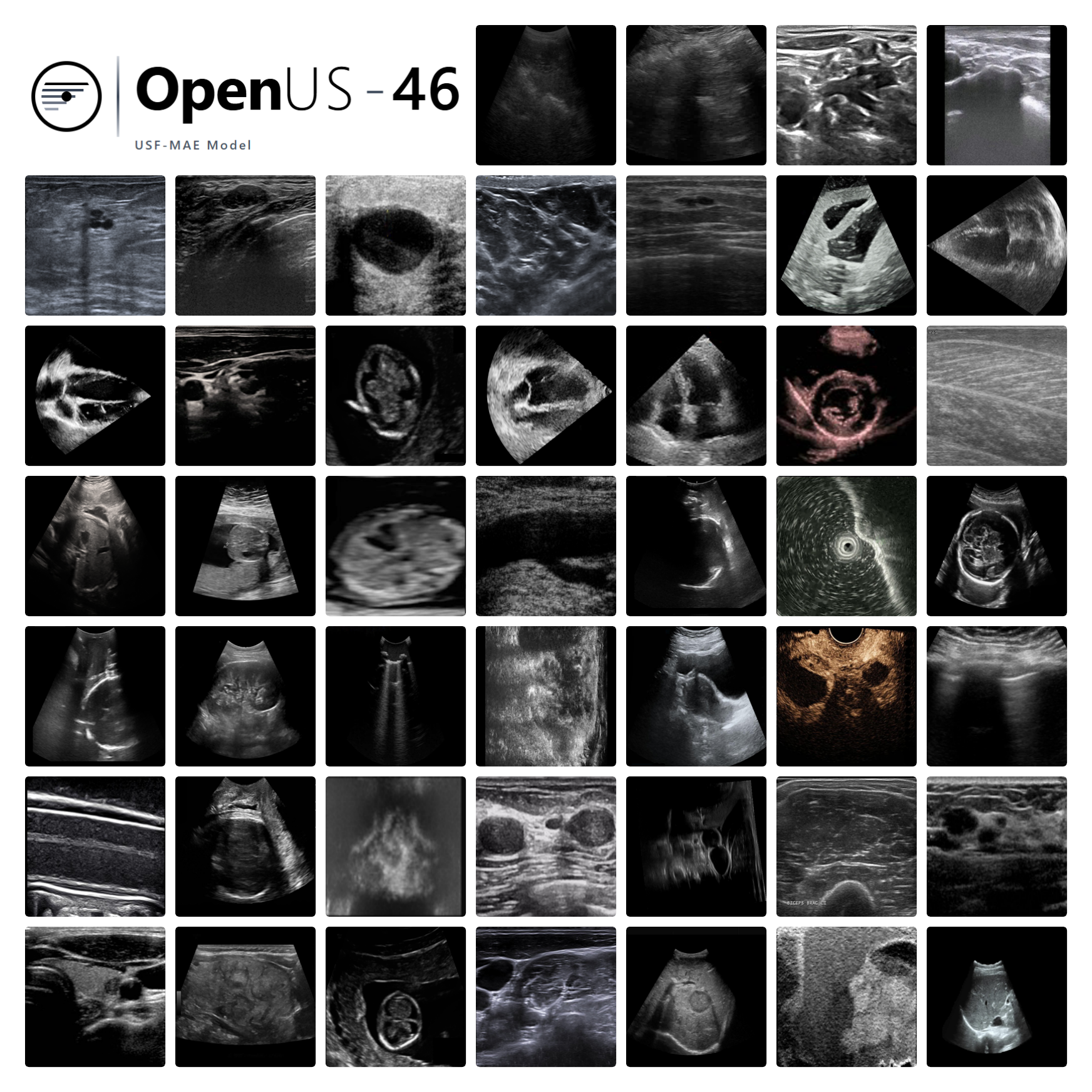}
	\caption{\textbf{OpenUS-46:} A curated collection of 46 open-source US images spanning diverse anatomical regions and clinical applications.}
	\label{FIG:OpenUS-46_figma}
\end{figure*}

Simultaneously, the domain of computer vision has seen the emergence of Vision Transformers (ViT) and associated self-attention models, which provide benefits in capturing long-range relationships inside images{~\hypersetup{hidelinks}\textcolor{blue}{\cite{b13}}}. While CNNs have been used as the foundation for medical image processing, including the majority of recent US AI applications, their confined receptive fields may restrict the acquisition of global context. Transformer architectures, first popularized in natural language processing, use self-attention to assess interactions among distant image patches, therefore capturing extensive contextual information. This skill is especially relevant for US, as differentiating actual structures from speckle noise or minor deviations necessitates the analysis of patterns over the whole image. Recent studies have shown that transformer-based models can surpass conventional CNNs in US applications. For example, integrating self-attention mechanisms into segmentation networks significantly enhanced accuracy in fetal US image segmentation relative to models using simply CNNs{~\hypersetup{hidelinks}\textcolor{blue}{\cite{b8}}}. The capacity to evaluate long-range spatial correlations enhances transformers' proficiency in managing US's speckle and diverse textures{~\hypersetup{hidelinks}\textcolor{blue}{\cite{b8}}}. Despite these advantages, transformer models have seen limited adoption in US to date. Transformers typically require very large training datasets to realize their full potential, and until recently, such datasets were not available in the US domain. Consequently, most prior US AI works continued to rely on CNN variants or hybrid CNN-Recurrent Neural Network (RNN) models, and the benefits of pure transformer architectures remained underexplored in this field.

The confluence of challenges in US imaging, noisy, variable data, and scarce labels, points to a need for approaches that learn generalizable representations without extensive annotations. This has led to growing interest in self-supervised learning and foundation models for medical images. Foundation models are large models pre-trained on vast amounts of data (often in a self-supervised or weakly-supervised manner) that can be adapted to a variety of downstream tasks{~\hypersetup{hidelinks}\textcolor{blue}{\cite{b14}}}. In medical imaging, such models promise to address the limitations of limited labeled data and improve cross-task generalization. Efforts to develop foundation models for US have only begun to emerge recently. These studies suggest that transformer-based foundation models can substantially improve US image analysis by learning domain-specific representations that transfer well to downstream problems.

A key bottleneck for US foundation models is the lack of organized, large-scale datasets. Publicly available US data tends to be fragmented into many small, task-specific sets (often focused on single organs or pathologies) that are collected with different protocols. This fragmentation makes it difficult to simply “train on all US data at once” as is done in natural image domains. Some recent works have confronted this issue by aggregating multiple datasets. Meyer et al. compiled US-43d, a collection of 43 open-access US datasets ($\sim$280,000 segmentation images with masks), spanning 20 clinical applications, to train their UltraSam model{~\hypersetup{hidelinks}\textcolor{blue}{\cite{b1}}}. This data-centric approach highlights that bringing together diverse ultrasound data from many sources is crucial to capture the modality’s inherent variability. In our work, we follow a similar philosophy by curating a large, unified ultrasound dataset from numerous open sources. Specifically, we gathered approximately 370,000 ultrasound images from 46 public datasets (\textbf{OpenUS-46}), which serves as an extension and refinement of the US-43d collection{~\hypersetup{hidelinks}\textcolor{blue}{\cite{b1}}}. Of the datasets originally included in US-43d, 36 were incorporated into OpenUS-46, while the remaining 7 datasets were excluded because certain datasets were no longer publicly accessible or had restricted/denied access, contained redundant entries, or represented duplicate versions of the same source under different names. These refinements were made solely to ensure dataset integrity and accessibility, while preserving the original intent of broad ultrasound coverage introduced by US-43d. {\hypersetup{hidelinks}\textcolor{blue}{Fig.~\ref{FIG:OpenUS-46_figma}}} shows one representative image from each dataset in OpenUS-46, covering a wide range of anatomical targets and clinical applications, forming a comprehensive pretraining corpus. Leveraging this corpus, we introduce \textbf{USF-MAE (Ultrasound Self-Supervised Foundation with Masked Autoencoding)} model (shown in {\hypersetup{hidelinks}\textcolor{blue}{Fig.~\ref{FIG:USF-MAE}}}), a transformer-based foundation model which is, to the best of our knowledge, the first large-scale self-supervised MAE in the US domain. USF-MAE is pre-trained to reconstruct masked US images, thereby learning rich latent representations of US structures without any manual labels.

While UltraSam{~\hypersetup{hidelinks}\textcolor{blue}{\cite{b1}}} demonstrated strong segmentation capabilities on the curated US-43d dataset, it is limited by reliance on large-scale manual annotations and a segmentation-specific design. In contrast, USF-MAE leverages $\sim$370,000 images from the OpenUS-46 ({\hypersetup{hidelinks}\textcolor{blue}{Fig.~\ref{FIG:OpenUS-46_figma}}}) in a self-supervised transformer framework, enabling task-agnostic pretraining without the use of any labels. We evaluate USF-MAE across diverse downstream classification tasks (breast lesion classification{~\hypersetup{hidelinks}\textcolor{blue}{\cite{b15}}}, gastrointestinal stromal tumor detection{~\hypersetup{hidelinks}\textcolor{blue}{\cite{b16}}}, and ovarian tumor subtype classification{~\hypersetup{hidelinks}\textcolor{blue}{\cite{b17}}}). USF-MAE consistently outperformed CNN and ViT baselines, and achieved performance competitive with or exceeding UltraSam. These results underscore that self-supervised, domain-specific pretraining yields broadly transferable US representations beyond segmentation.

\begin{figure*}
	\centering
	\includegraphics[width=.9\textwidth]{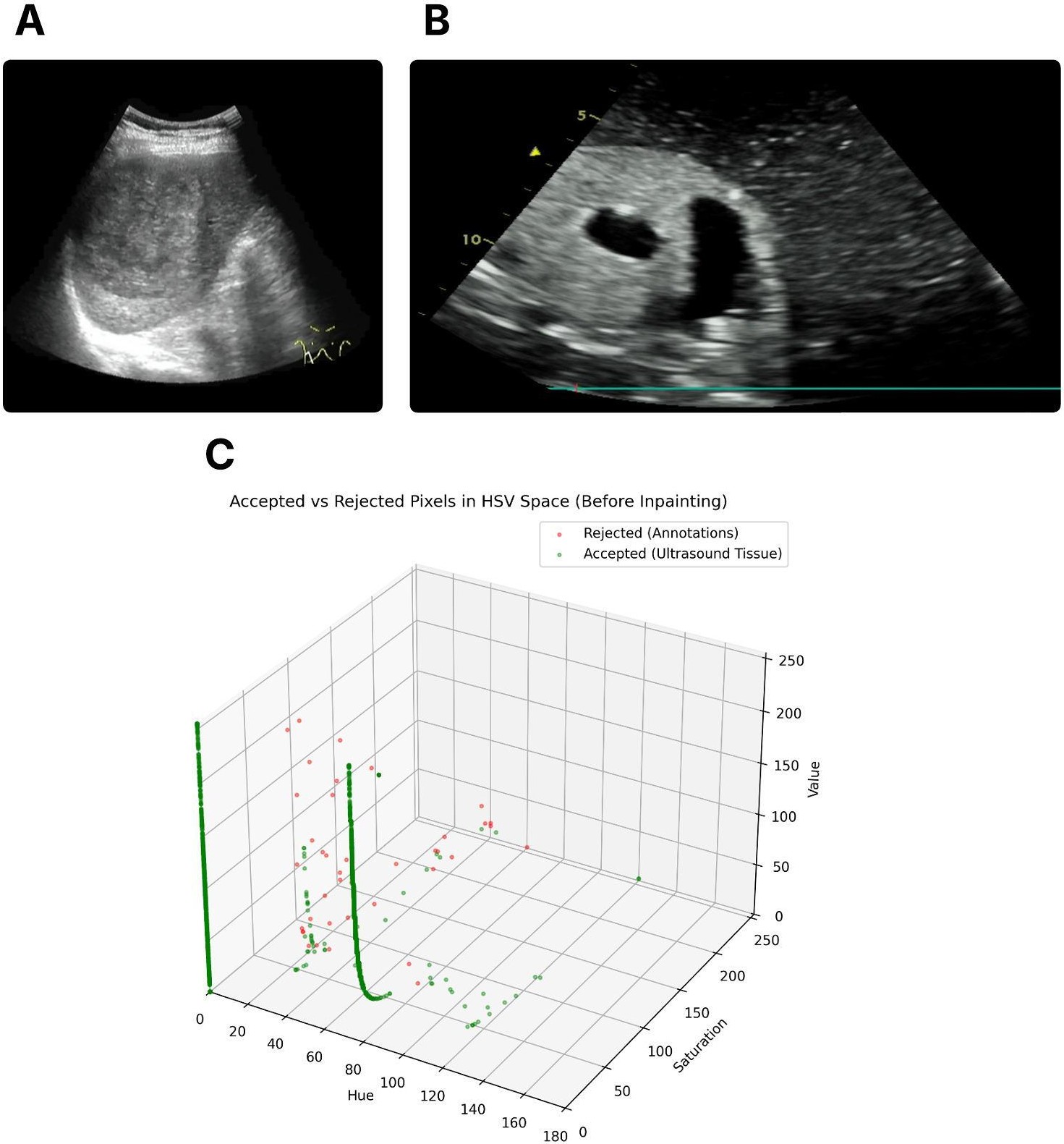}
	\caption{US images of two scans from AUL dataset(A) and Cactus dataset (B), and visualization of accepted vs. rejected pixels in HSV color space before inpainting (C). Accepted pixels correspond to US tissue, while rejected pixels represent annotation overlays.}
	\label{FIG:hsv_preprcoessing_figure}
\end{figure*}

\section{Methodology}
\subsection{Dataset Curation}
We curated OpenUS-46 ({\hypersetup{hidelinks}\textcolor{blue}{Fig.~\ref{FIG:OpenUS-46_figma}}}), comprising 46 publicly available US datasets, covering 23 distinct clinical applications and totaling approximately 370,000 2D and 3D scans{~\hypersetup{hidelinks}\textcolor{blue}{\cite{b28,b29,b30,b31,b32,b33,b34,b35,b36,b37,b38,b39,b40,b41,b42,b44,b45,b46,b47,b48,b49,b50,b51,b52,b53,b54,b55,b57,b58,b59,b60,b62,b63,b64,b65,b66,b67,b68,b69,b71,b74,b75}}}. An additional 10,578 images were reserved for evaluation{~\hypersetup{hidelinks}\textcolor{blue}{\cite{b15,b16,b17}}}. The datasets were sourced from multiple open platforms, including Papers with Code, Google Dataset Search, GitHub, Kaggle, ResearchGate, Mendeley Data, and Zenodo. Furthermore, contributions were included from institutional releases by the University of Electronic Science and Technology of China, Stanford University Medical Center, the Technical University of Munich, and other centers. 

The aggregated dataset captures organs and lesions of various shapes, sizes, and textures across applications such as pelvis, fetal head, thyroid nodules, breast lesions, gastrointestinal tumors, kidney, lung, head and neck, and musculoskeletal structures, thereby providing a comprehensive foundation for generalizable US representation learning. A detailed summary of the datasets, including their clinical focus and image counts, is provided in{~\hypersetup{hidelinks}\textcolor{blue}{Table~\ref{tab:us_datasets_summary}}} while the dataset links are available on our \href{https://github.com/Yusufii9/USF-MAE}{\textcolor{blue}{\textbf{GitHub Repository}}}\\(https://github.com/Yusufii9/USF-MAE).

Some datasets in OpenUS-46 consisted of US \textit{video sequences} rather than static images (e.g., echocardiography and fetal motion datasets). To ensure balanced representation without redundancy, frames were uniformly sampled at a rate of three frames per second. This sampling strategy prevents over-representation of nearly identical frames and promotes diversity in anatomical views and probe orientations. Extracted frames were then treated as independent 2D images and incorporated into the final dataset collection, consisting of 374,192 images exactly, for preprocessing and pretraining.

\subsection{Preprocessing Pipeline}
The collected datasets exhibit significant heterogeneity, not only in terms of acquisition parameters but also in terms of overlays and inconsistent image shapes. To address these issues, we developed a multi-stage preprocessing pipeline that systematically normalizes all images before pretraining.

\begin{figure*}
	\centering
	\includegraphics[width=.98\textwidth]{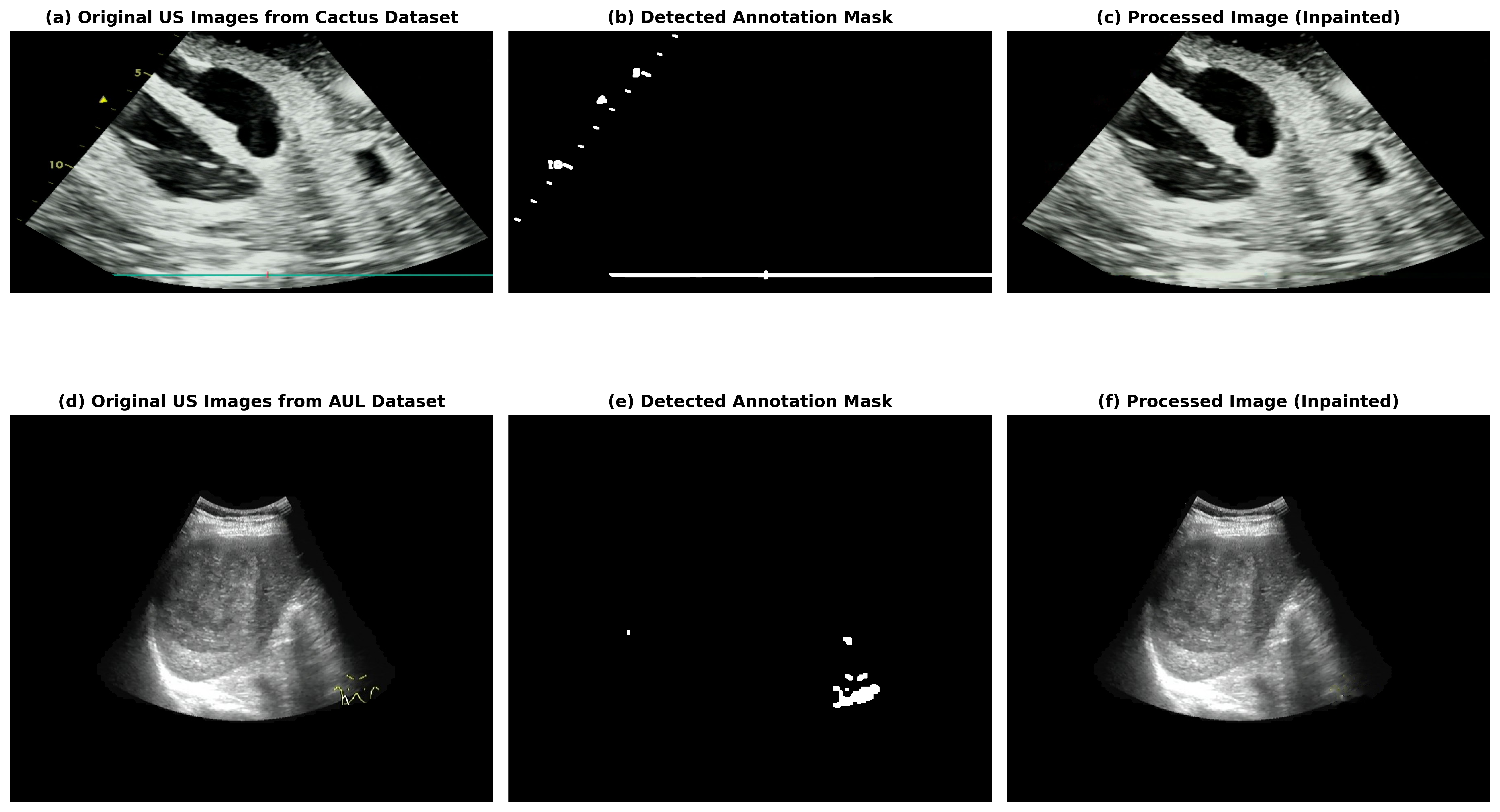}
	\caption{US image preprocessing pipelines used for annotation removal and standardization. (a-c) Cactus dataset workflow: original US image, automatically detected colored annotation mask, and the corresponding inpainted (cleaned) image using the Navier-Stokes algorithm{~\textcolor{blue}{[67]}}. (d-f) AUL dataset workflow: original US image, detected annotation mask using CLAHE-enhanced K-means clustering, and the inpainted result.}
	\label{FIG:Ultrasound_Preprocessing_Pipelines}
\end{figure*}

As a first step, these annotations were removed because many datasets contain overlaid text or graphical markers, such as labels, arrows, lesion outlines, and measurements, shown in {\hypersetup{hidelinks}\textcolor{blue}{Fig.~\ref{FIG:hsv_preprcoessing_figure}}}. We implemented a unified detection and inpainting framework to automatically identify and suppress these annotations. Text elements were localized in grayscale images using Tesseract OCR (Optical Character Recognition), and bounding boxes were converted into binary masks. To detect weak or low-contrast annotations, contrast-limited adaptive histogram equalization (CLAHE) was applied to the luminance channel of the CIELAB colour space to improve visibility before detection. Colour-based annotations (including highlighter strokes and ink) were segmented in the HSV colour space, with {\hypersetup{hidelinks}\textcolor{blue}{Fig.~\ref{FIG:hsv_preprcoessing_figure}C}} showing a 3D visualization of pixel distribution in HSV space, and highly saturated pixels are clustered using K-Means clustering. Grayscale annotations (such as pencil marks and underlines) were separated using Canny edge detection, followed by contour extraction. Masks generated from the text, colour, and grayscale channels were combined into a unified annotation mask, which was then refined using morphological closing and opening operations. Finally, a Navier-Stokes-based{~\hypersetup{hidelinks}\textcolor{blue}{\cite{b72}}} inpainting algorithm (implemented in OpenCV) was applied to reconstruct the background structure within the masked region. This procedure effectively removed diverse annotation types while preserving underlying US information ({\hypersetup{hidelinks}\textcolor{blue}{Fig.~\ref{FIG:Ultrasound_Preprocessing_Pipelines}}} - example of annotation removal).

Following annotation removal, we applied normalization to harmonize image intensity distributions. All static and video-derived frame scans were normalized according to common specifications for computer vision models, with a mean of [0.485, 0.456, 0.406] and a standard deviation of [0.229, 0.224, 0.225]. This ensures consistent pixel intensity across all datasets and enables stable convergence during pre-training.

Lastly, the spatial dimensions of the images were normalized. All images were resized to 224$\times$224 pixels, a widely used resolution in Transformer-based architectures. This step ensures compatibility with common pre-trained backbones used for benchmarking experiments.

The entire preprocessing pipeline ran in batch mode, and all processed outputs were saved in PNG format. This approach thoroughly normalized the selected collection of $\sim$370,000 US images, making them appropriate for pre-training the USF-MAE model.

\begin{figure*}
	\centering
	\includegraphics[width=1\textwidth]{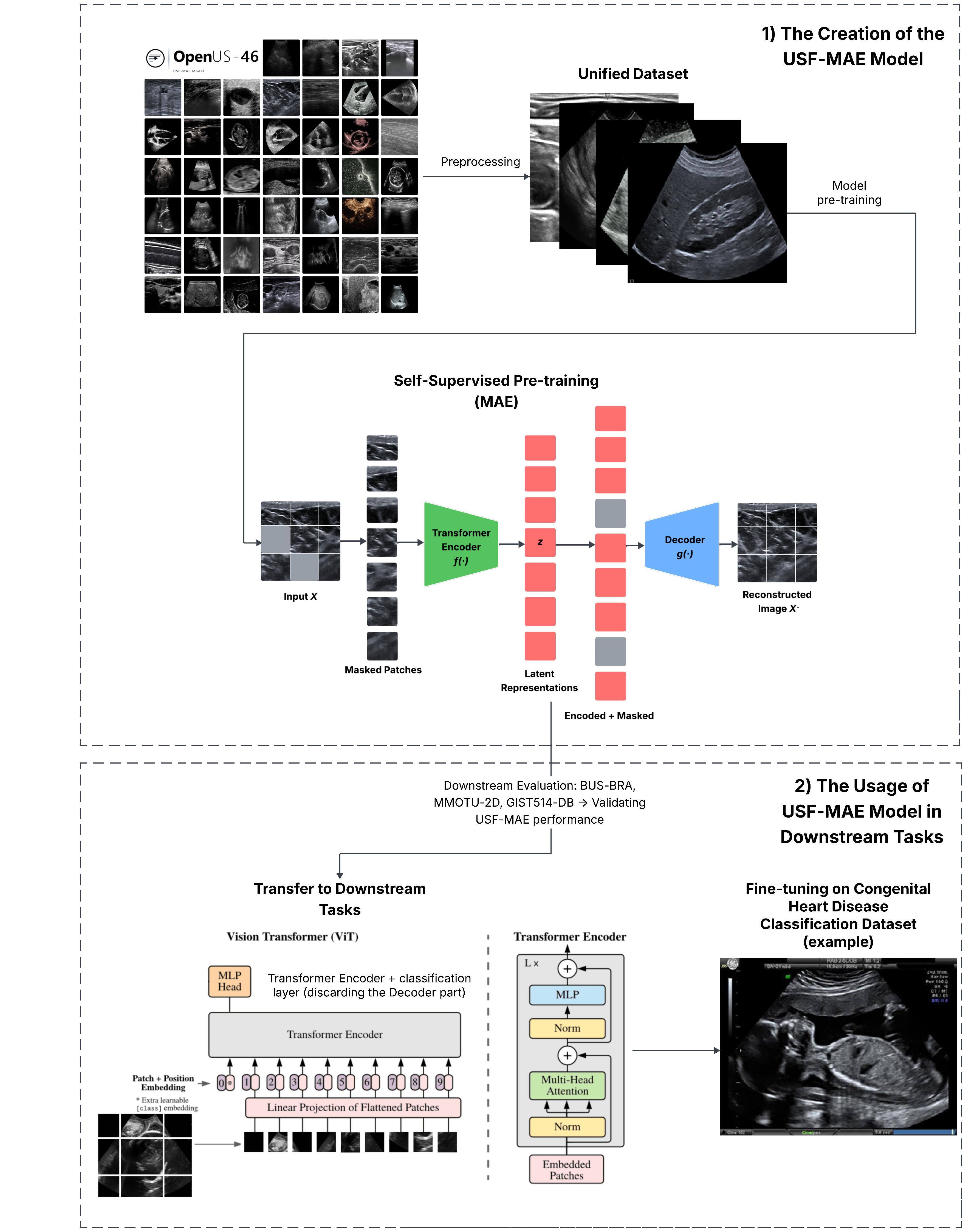}
	\caption{Overview of the USF-MAE (Ultrasound Self-Supervised Foundation with Masked Autoencoding) pipeline for 1) pre-training and 2) downstream fine-tuning.}
	\label{FIG:USF-MAE}
\end{figure*}

\subsection{USF-MAE Architecture}
The proposed USF-MAE (shown in{~\hypersetup{hidelinks}\textcolor{blue}{Fig.~\ref{FIG:USF-MAE}}}) is based on the MAE framework{~\hypersetup{hidelinks}\textcolor{blue}{\cite{b73}}}, which was modified to improve performance for US imaging. The model uses a ViT backbone as an encoder and a lightweight transformer decoder for image reconstruction. The model can acquire strong and generalizable representations of US data via self-supervised pretraining following this approach.

As illustrated in {\hypersetup{hidelinks}\textcolor{blue}{Fig.~\ref{FIG:USF-MAE}}}, the architecture follows a masked autoencoding strategy, where US images are partitioned into patches, partially masked, encoded with a ViT backbone, and reconstructed through a transformer decoder. 

% \noindent\textbf{Patch Embedding and Encoding.} 
% Each input image $x \in \mathbb{R}^{H \times W \times C}$ is divided into 
% $N$ patches $\{x_i\}_{i=1}^{N}$, each flattened and linearly projected as:
% \begin{equation}
% z_i^0 = E x_i + p_i,
% \end{equation}
% where $E$ is the learnable embedding matrix and $p_i$ is the positional encoding for patch $i$. 
% The encoder then applies $L$ transformer layers, where each layer updates representations as:
% \begin{equation}
% z^{l+1} = \text{MLP}\bigl(\text{LN}(z^l + \text{MSA}(\text{LN}(z^l)))\bigr),
% \quad l = 1, \dots, L,
% \end{equation}
% with $\text{MSA}$ denoting multi-head self-attention and $\text{LN}$ layer normalization.

% 1st new edit
\noindent\textbf{Patch Embedding and Encoding.}
Each input image $x \in \mathbb{R}^{H \times W \times C}$ is divided into 
$N$ non-overlapping patches $\{x_i\}_{i=1}^{N}$, each of size $P \times P$ 
pixels. Each flattened patch is linearly projected into a latent embedding 
space and augmented with a learnable positional encoding as:

\begin{equation}
z_i^0 = E x_i + p_i,
\label{eq:patch_embed}
\end{equation}

\noindent
where $E \in \mathbb{R}^{d \times P^2C}$ is the patch embedding matrix and 
$p_i$ is the positional encoding vector for patch $i$. The encoder processes 
these embeddings through $L$ stacked transformer layers. Each layer updates 
the token representations via multi-head self-attention (MSA) and a feed-forward 
network (MLP):

\begin{equation}
z^{l+1} = \text{MLP}\bigl(\text{LN}(z^l + \text{MSA}(\text{LN}(z^l)))\bigr),
\quad l = 1, \dots, L,
\label{eq:vit_layer}
\end{equation}

\noindent
where $\text{LN}(\cdot)$ denotes layer normalization. The output of the final 
encoder layer, $z^L$, provides compact and context-rich latent representations 
of the visible ultrasound patches.

The encoder-decoder structure followed the principle of reconstructing missing visual information from incomplete observations. Each input US image was first divided into non-overlapping patches of fixed size. In our implementation, images of resolution 224$\times$224 were partitioned into patches of 16$\times$16 pixels, resulting in 196 patches per image. A fixed proportion of these patches, 25\% in our case, was randomly masked using a uniform sampling strategy without replacement{~\hypersetup{hidelinks}\textcolor{blue}{\cite{b73}}}. The remaining 75\% of patches, which constitute the visible input, were linearly projected into an embedding space and supplemented with learnable positional encodings. These embedded tokens were then processed by the ViT encoder, which comprised 12 transformer layers, each with a hidden embedding dimension of 768 and 12 self-attention heads. The encoder produced latent representations that summarize the visible content of the input image.

To enable image reconstruction, the decoder received a full set of tokens consisting of the latent representations of visible patches and placeholder tokens corresponding to the masked patches{~\hypersetup{hidelinks}\textcolor{blue}{\cite{b73}}}. Each mask token was represented by a learnable vector that signifies the absence of visual input at a particular spatial location. Positional embeddings were again added to the full token sequence to preserve spatial alignment. The decoder itself consisted of 8 transformer layers with an embedding dimension of 512 and 8 attention heads. Its sole purpose was to predict pixel intensities for the masked patches, thereby reconstructing the full image. Importantly, the decoder was used only during pretraining. Once pretraining is complete, the encoder serves as the foundation model to generate image representations for downstream tasks, while the decoder is discarded.

This design ensures that USF-MAE learns compact and context-rich latent features by focusing its capacity on modeling the visible context and inferring the missing structure of US images. Such representations, learned in a self-supervised manner from hundreds of thousands of diverse US scans, provide a powerful initialization for fine-tuning across a variety of diagnostic tasks.

\subsection{Pretraining Setup}
The USF-MAE model was pretrained in a self-supervised fashion using a masked reconstruction objective. In the MAE architecture, input US images were divided into non-overlapping 16$\times$16 patches and randomly masked with a ratio of 25\%. Visible patches were then used to train an encoder that will produce latent representations, while a lightweight decoder reconstructs the masked patches. Training loss is thus defined as the Mean Squared Error (MSE) between reconstructed pixels and original pixels over just the masked regions. \noindent\textbf{Loss Function.}  % 2nd new edit
Formally, let an input US image be denoted as 
$x \in \mathbb{R}^{H \times W \times C}$, which is divided into 
$N$ non-overlapping patches $\{x_i\}_{i=1}^{N}$ of size 
$P \times P$, where each patch has $d = P^2 C$ pixels. 
Let $\mathcal{M} \subset \{1, \dots, N\}$ denote the index set of masked patches 
(with $|\mathcal{M}| = M$), and let $\hat{x}_i$ represent the reconstructed patch 
corresponding to $x_i$. The reconstruction objective is computed only over the masked 
patches using the pixel-wise MSE:

\begin{equation}
\mathcal{L}_{\text{MAE}} 
= \frac{1}{M d} \sum_{i \in \mathcal{M}} 
\bigl\| \hat{x}_i - x_i \bigr\|_2^2,
\label{eq:mae_loss}
\end{equation}

\noindent
where $M$ is the number of masked patches and $d$ is the number of pixels per patch. 
Equivalently, in matrix form:

\begin{equation}
\mathcal{L}_{\text{MAE}} 
= \frac{1}{M d} \bigl\| (\hat{X} - X) \odot M \bigr\|_F^2,
\label{eq:mae_loss_matrix}
\end{equation}

\noindent
where $X, \hat{X} \in \mathbb{R}^{N \times d}$ stack the original and reconstructed 
patches, respectively, $\odot$ denotes element-wise multiplication, 
and $\mathbf{M} \in \{0,1\}^{N \times d}$ is the binary masking matrix. 

The gradient with respect to each reconstructed patch is given by:

\begin{equation}
\frac{\partial \mathcal{L}_{\text{MAE}}}{\partial \hat{x}_i} =
\begin{cases}
\displaystyle \frac{2}{M d} (\hat{x}_i - x_i), & i \in \mathcal{M}, \\[6pt]
0, & i \notin \mathcal{M}.
\end{cases}
\label{eq:mae_grad}
\end{equation}

\noindent
This BERT-style masking strategy ensures that only masked patches contribute to the 
loss, compelling the encoder to infer the missing ultrasound structures from visible 
context and thereby learn context-aware, noise-tolerant representations.

% This objective pushes the network to learn informative features about anatomical structures and texture content features important in US, since its images are corrupted by speckle noise and varying acquisition conditions.

To optimize the model, we used the AdamW optimizer with an initial learning rate of 0.001 and a weight decay of 0.01. These hyperparameters were selected after performing 5-fold cross-validation, where combinations of learning rates (0.0003 and 0.001) and weight decays (0.01 and 0.05) were compared, and the chosen values corresponded to the lowest average validation loss across folds. The nominal learning rate followed scaling by batch size (set to 64) as per common practice for transformer pretraining. The cosine schedule for the learning rate with warmup was used such that it attained its peak value linearly in the first 10\% of steps and then decayed smoothly according to a half-cycle cosine schedule. Training stability was maintained through gradient clipping at a maximum norm of 1.0. In addition, we applied data augmentations during pretraining to enhance generalization while preserving US texture integrity. These included random rotations between 0\textdegree \text{ and} 90\textdegree, horizontal and vertical flips (each with a probability of 0.5), and random resized cropping with scaling between 0.5$\times$ and 2.0$\times$ of the original resolution.

Pre-training took place on a machine powered by an NVIDIA RTX 4080 Super GPU and 96 GB system RAM. Training for 100 epochs consumed approximately 26.5 hours, whereas pushing the training up to 500 epochs took 131 hours. Empirically, we found that the model trained for 100 epochs already provided strong performance across downstream tasks, and this version is used for all evaluations presented in this work. However, to promote reproducibility and exploration by the research community, we make available both the 100-epoch and 500-epoch pretrained weights.

The chosen hyperparameters were selected to balance computational efficiency with representational quality. A moderate masking ratio of 25\% was used after preliminary experiments showed it provided better downstream transferability than higher masking ratios (e.g., 75-90\%), especially with medical images{~\hypersetup{hidelinks}\textcolor{blue}{\cite{b21}}}, which tended to degrade fine-grained anatomical detail in reconstructed images. Similarly, the 100-epoch schedule was sufficient for convergence given the large and diverse dataset, whereas training to 500 epochs yielded similar performance at the cost of a significant increase in compute time.

\subsection{Downstream Tasks}
To assess the generalizability of the USF-MAE model across diverse diagnostic environments, we fine-tuned the pre-trained encoder on three publicly accessible downstream US datasets following the bottom block shown in{~\hypersetup{hidelinks}\textcolor{blue}{Fig.~\ref{FIG:USF-MAE}}} pipeline. Importantly, none of these datasets were included in the pretraining corpus, ensuring a fair and unbiased evaluation of the model’s transferability to unseen data. Together, these three downstream tasks span both binary and multi-class classification, cover a variety of anatomical regions (breast, ovary, intestinal tract), and have a significant clinical application. They provide a comprehensive evaluation system that assesses the degree to which our model is generalizable across tasks. In addition, these three datasets (BUS-BRA{~\hypersetup{hidelinks}\textcolor{blue}{\cite{b15}}}, MMOTU-2D{~\hypersetup{hidelinks}\textcolor{blue}{\cite{b17}}}, and GIST514-DB{~\hypersetup{hidelinks}\textcolor{blue}{\cite{b16}}}) were employed in the UltraSam paper{~\hypersetup{hidelinks}\textcolor{blue}{\cite{b1}}}, allowing us to directly compare the performance of USF-MAE against this prior foundation model.

% 3rd new edit
\noindent\textbf{Fine-Tuning Objective.}
During supervised fine-tuning, the pretrained encoder was coupled with a classification 
head and optimized using the standard cross-entropy loss, as implemented in 
\texttt{torch.nn.CrossEntropyLoss()}. Given the predicted logits 
$z_i \in \mathbb{R}^{C}$ for each sample $i$ and its ground-truth class label 
$y_i \in \{1, \dots, C\}$, the loss is defined as:

\begin{equation}
\mathcal{L}_{\text{CE}} 
= -\frac{1}{N} \sum_{i=1}^{N} 
\log \frac{\exp(z_{i, y_i})}
{\sum_{c=1}^{C} \exp(z_{i, c})},
\label{eq:cross_entropy_pytorch}
\end{equation}

\noindent
where $N$ is the batch size and $C$ is the total number of classes. 
This formulation corresponds to the negative log-likelihood of the true class 
under the model’s softmax output, and it complements the self-supervised 
reconstruction objective {\hypersetup{hidelinks}\textcolor{blue}{Eq.~\ref{eq:mae_loss}}} used during pretraining.

\begin{figure}
	\centering
	\includegraphics[width=.5\textwidth]{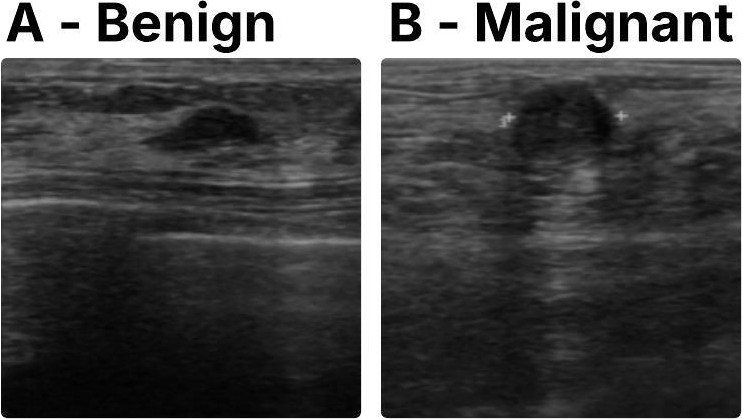}
	\caption{US images of Benign (A) and Malignant (B) scans from the BUS-BRA dataset.}
	\label{FIG:BUS-BRA_examples}
\end{figure}

\textbf{1) Breast Cancer (BUS-BRA).}
The BUS-BRA set{~\hypersetup{hidelinks}\textcolor{blue}{\cite{b15}}} has 1875 breast US images taken from 1064 women patients marked as either benign (n=722) or malignant (n=342) cases for binary classification, an example of each class is shown in{~\hypersetup{hidelinks}\textcolor{blue}{Fig.~\ref{FIG:BUS-BRA_examples}}}. We used the official split in 5-fold cross-validation given by the dataset's authors{~\hypersetup{hidelinks}\textcolor{blue}{\cite{b15}}}. Breast cancer happens to be the top cancer among females globally{~\hypersetup{hidelinks}\textcolor{blue}{\cite{b15}}}. US is a key method in finding suspicious spots, mainly in thick breast tissue. This dataset presents a tough challenge because of the fine visual gaps between benign and malignant lesions and speckle noise's commonness in breast US images.

\begin{figure}
	\centering
	\includegraphics[width=.48\textwidth]{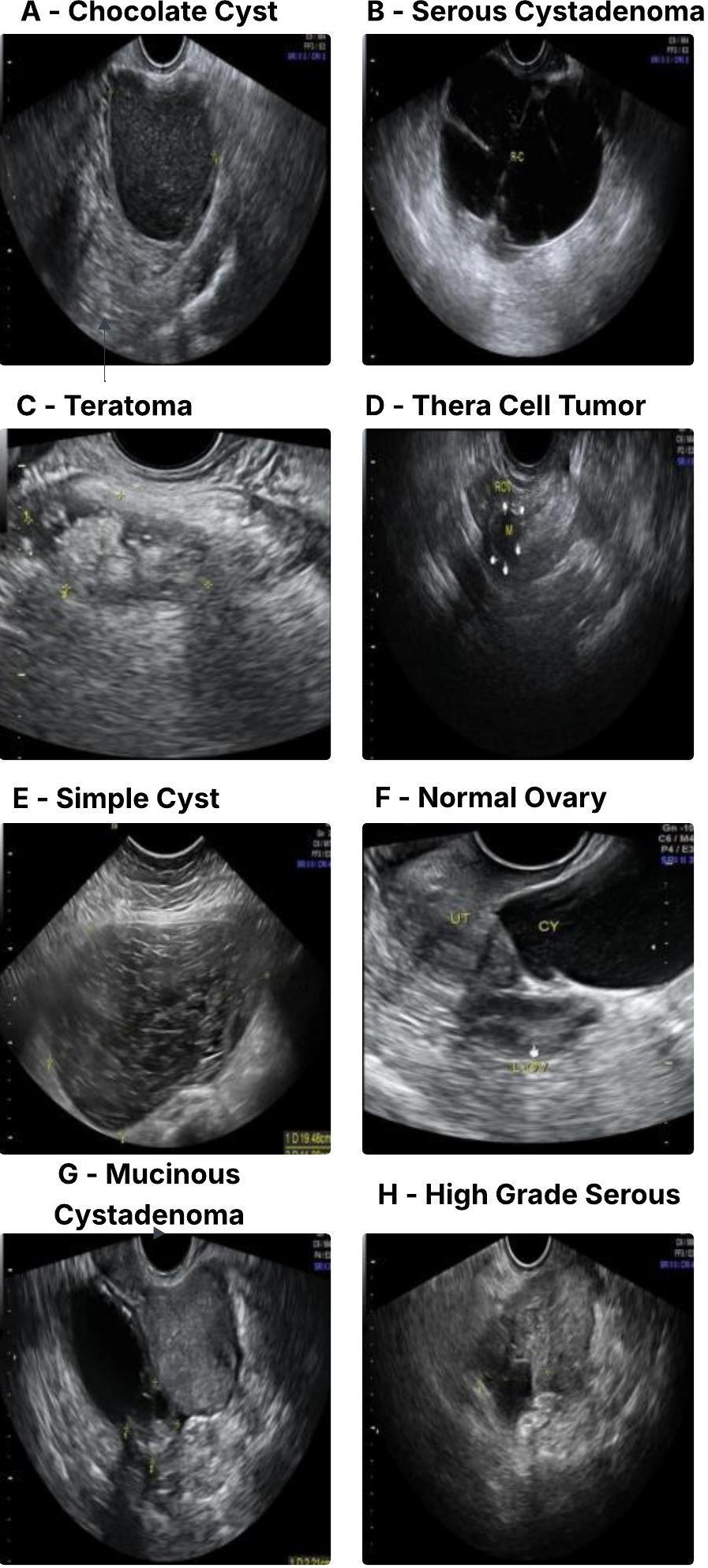}
	\caption{US images of various ovarian lesions, including entities such as Chocolate Cyst (A), Serous Cystadenoma (B), Teratoma (C), Thera Cell Tumor (D), Simple Cyst (E), Normal Ovary (F), Mucinous Cystadenoma (G), and High Grade Serous (H) from the MMOTU-2D dataset.}
	\label{FIG:MMOTU_examples}
\end{figure}

\textbf{2) Ovarian Tumors (MMOTU-2D).}
The MMOTU-2D dataset{~\hypersetup{hidelinks}\textcolor{blue}{\cite{b17}}} consists of 1469 2D ovarian tumor US images from 247 patients, annotated into eight distinct pathological subtypes (chocolate cyst, serous cystadenoma, teratoma, thera cell tumour, simple cyst, normal ovary, mucinous cystadenoma, high grade serous), as seen in {\hypersetup{hidelinks}\textcolor{blue}{Fig.~\ref{FIG:MMOTU_examples}}}. We followed the official split of 1000 images for training and 469 for validation provided by the dataset authors{~\hypersetup{hidelinks}\textcolor{blue}{\cite{b17}}}. The multi-class classification scheme on this dataset evaluates the capacity of the USF-MAE model to differentiate among multiple tumour types that have different appearances. The dataset's clinical application is significant, as accurate identification of the ovarian tumor subtypes is crucial to the planning of treatments. This dataset is a valuable resource for evaluating the performance of specific subtypes of ovarian tumours.

\begin{figure}
	\centering
	\includegraphics[width=.5\textwidth]{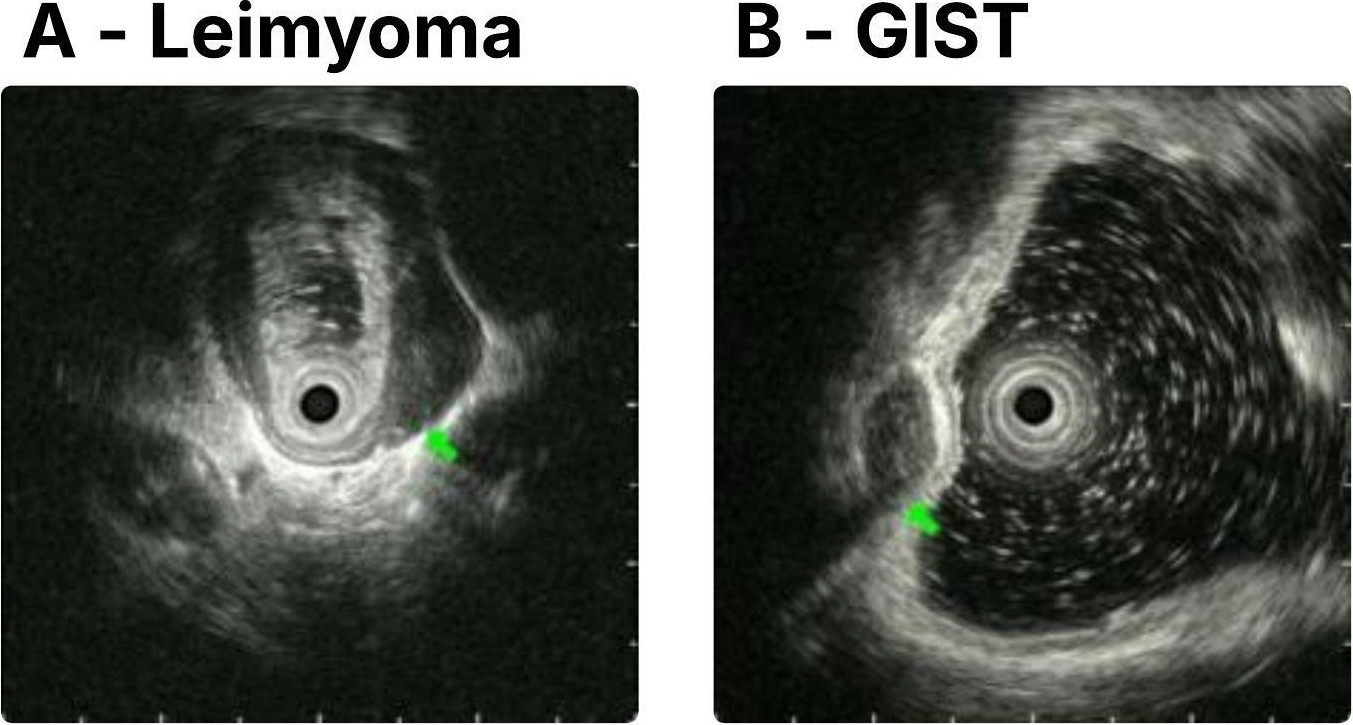}
	\caption{US images of Leimyoma (Leiomyoma) (A) and Gastrointestinal Stromal Tumor (GIST) (B) scans from the GIST514-DB dataset.}
	\label{FIG:GIST_examples}
\end{figure}

\textbf{3) Gastrointestinal Stromal Tumor (GIST514-DB).}
The GIST514-DB dataset{~\hypersetup{hidelinks}\textcolor{blue}{\cite{b16}}} contains 514 US images that are classified as either 251 gastrointestinal stromal tumours (GISTs) or 263 cases of leiomyoma (an example of each type can be seen in {\hypersetup{hidelinks}\textcolor{blue}{Fig.~\ref{FIG:GIST_examples}}}). GISTs are the most common types of mesenchymal tumours in the stomach's tract, and their distinction from non-neoplastic conditions is essential guidance for clinical management. The dataset is comprised of carefully annotated US images, which provide a means of comparison to assess the performance of USF-MAE in the detection of gastrointestinal tumours in real-world conditions.

\begin{table*}[htbp]
\centering
\caption{Summary of collected US datasets (OpenUS-46), their target anatomies, and image counts. Datasets in \textbf{bold} are also included in US-43d~\textcolor{blue}{\cite{b1}}.}

% \caption{Summary of collected US datasets (OpenUS-46), their target anatomies, and image counts.}
\label{tab:us_datasets_summary}
\setlength{\tabcolsep}{10pt} % Adjust column spacing if needed
\begin{tabular}{lcc}
\toprule
\textbf{Dataset Name} & \textbf{Target Anatomy} & \textbf{\# Images Used} \\
\midrule
\textbf{105US}{~\hypersetup{hidelinks}\textcolor{blue}{\cite{b33}}} & Liver & 105  \\
\textbf{AbdomenUS}{~\hypersetup{hidelinks}\textcolor{blue}{\cite{b34}}} & Abdomen & 1544 \\
\textbf{ACOUSLIC}{~\hypersetup{hidelinks}\textcolor{blue}{\cite{b35}}} & Abdomen & 300 \\
\textbf{AUL}{~\hypersetup{hidelinks}\textcolor{blue}{\cite{b36}}} & Liver & 735 \\
BLUES{~\hypersetup{hidelinks}\textcolor{blue}{\cite{b37}}} & Lung & 4133 \\
\textbf{BP}{~\hypersetup{hidelinks}\textcolor{blue}{\cite{b39}}} & Neck & 11143 \\
\textbf{brachial plexus}{~\hypersetup{hidelinks}\textcolor{blue}{\cite{b38}}} & Neck & 7462 \\
\textbf{BrEaST}{~\hypersetup{hidelinks}\textcolor{blue}{\cite{b40}}} & Breast & 256 \\
\textbf{BUS (Dataset B)}{~\hypersetup{hidelinks}\textcolor{blue}{\cite{b44}}} & Breast & 163 \\
\textbf{BUS\_UC}{~\hypersetup{hidelinks}\textcolor{blue}{\cite{b41}}} & Breast & 811 \\
\textbf{BUS\_UCML}{~\hypersetup{hidelinks}\textcolor{blue}{\cite{b42}}} & Breast & 683 \\
\textbf{BUS-BRA}{~\hypersetup{hidelinks}\textcolor{blue}{\cite{b15}}} & Breast & 1875 \\
Cactus Dataset{~\hypersetup{hidelinks}\textcolor{blue}{\cite{b30}}} & Heart & 37736 \\
CAMUS\_public{~\hypersetup{hidelinks}\textcolor{blue}{\cite{b46}}} & Echocardiography & 19237 \\
\textbf{CardiacUDC}{~\hypersetup{hidelinks}\textcolor{blue}{\cite{b47}}} & Heart & 760 \\
\textbf{CCAUI}{~\hypersetup{hidelinks}\textcolor{blue}{\cite{b48}}} & Common Carotid Artery & 1100 \\
\textbf{DFHI}{~\hypersetup{hidelinks}\textcolor{blue}{\cite{b45}}} & Fetal Head & 3832 \\
\textbf{EchoCP}{~\hypersetup{hidelinks}\textcolor{blue}{\cite{b49}}} & Heart & 11519 \\
\textbf{EchoNet-Dynamic}{~\hypersetup{hidelinks}\textcolor{blue}{\cite{b50}}} & Heart & 112740 \\
\textbf{EchoNet-Pediatric}{~\hypersetup{hidelinks}\textcolor{blue}{\cite{b51}}} & Heart & 46573 \\
\textbf{FALLMUD}{~\hypersetup{hidelinks}\textcolor{blue}{\cite{b52}}} & Leg Muscle & 813 \\
\textbf{FASS}{~\hypersetup{hidelinks}\textcolor{blue}{\cite{b53}}} & Fetal Abdominal & 1588 \\
\textbf{Fast-U-Net}{~\hypersetup{hidelinks}\textcolor{blue}{\cite{b54}}} & Fetal Abdominal & 1414 \\
FEFT{~\hypersetup{hidelinks}\textcolor{blue}{\cite{b74,b75}}} & Fetal Echocardiography & 6720 \\
FETAL\_PLANES\_DB{~\hypersetup{hidelinks}\textcolor{blue}{\cite{b28}}} & Fetal Anatomy & 12400 \\
\textbf{FH-PS-AOP}{~\hypersetup{hidelinks}\textcolor{blue}{\cite{b55}}} & Fetal Head & 4000 \\
\textbf{GIST514-DB}{~\hypersetup{hidelinks}\textcolor{blue}{\cite{b16}}} & Gastrointestinal & 514 \\
\textbf{HC}{~\hypersetup{hidelinks}\textcolor{blue}{\cite{b57}}} & Fetal Head Circumference & 1334 \\
JNU-IFM{~\hypersetup{hidelinks}\textcolor{blue}{\cite{b32}}} & Pelvis and Fetal Head & 6224 \\
\textbf{kidneyUS}{~\hypersetup{hidelinks}\textcolor{blue}{\cite{b58}}} & Kidney & 534 \\
\textbf{LUSS\_phantom}{~\hypersetup{hidelinks}\textcolor{blue}{\cite{b59}}} & Lung & 564 \\
\textbf{MicroSeg}{~\hypersetup{hidelinks}\textcolor{blue}{\cite{b60}}} & Prostate & 2910 \\
\textbf{MMOTU-2D}{~\hypersetup{hidelinks}\textcolor{blue}{\cite{b17}}} & Ovarian Tumor & 1469 \\
\textbf{MMOTU-3D}{~\hypersetup{hidelinks}\textcolor{blue}{\cite{b17}}} & Ovarian Tumor & 170 \\
Pocus{~\hypersetup{hidelinks}\textcolor{blue}{\cite{b62}}} & Lung & 2264 \\
Porcine{~\hypersetup{hidelinks}\textcolor{blue}{\cite{b63}}} & Porcine Spinal Cord & 12468 \\
PSFHS{~\hypersetup{hidelinks}\textcolor{blue}{\cite{b29}}} & Pelvis and Fetal Head & 1358 \\
\textbf{regPro}{~\hypersetup{hidelinks}\textcolor{blue}{\cite{b64}}} & Prostate & 6424 \\
\textbf{S1}{~\hypersetup{hidelinks}\textcolor{blue}{\cite{b65}}} & Breast & 201 \\
\textbf{Segthy}{~\hypersetup{hidelinks}\textcolor{blue}{\cite{b66}}} & Thyroid and Neck & 15820 \\
\textbf{STMUS\_NDA}{~\hypersetup{hidelinks}\textcolor{blue}{\cite{b67}}} & Musculoskeletal & 8169 \\
\textbf{STU-Hospital}{~\hypersetup{hidelinks}\textcolor{blue}{\cite{b68}}} & N/A & 42 \\
\textbf{Thyroid Dataset}{~\hypersetup{hidelinks}\textcolor{blue}{\cite{b31}}} & Head and Neck & 3838 \\
\textbf{Thyroid US Cineclip}{~\hypersetup{hidelinks}\textcolor{blue}{\cite{b69}}} & Thyroid & 17412 \\
Ultrasound Fetus Dataset{~\hypersetup{hidelinks}\textcolor{blue}{\cite{b71}}} & Fetal & 1880 \\
\textbf{UPBD}{~\hypersetup{hidelinks}\textcolor{blue}{\cite{b33}}} & Brachial Plexus & 955 \\
\bottomrule
\end{tabular}
\end{table*}

\subsection{Baseline Models for Comparison}
\textbf{CNN-based Models (VGG-19 and ResNet-50)}. Convolutional Neural Networks (CNNs) have been the dominant architecture for image recognition for the past decade. We include two CNN baselines, VGG-19 and ResNet-50 , which are both widely used backbone models. \textbf{VGG-19} is a 19-layer network introduced by Simonyan and Zisserman{~\hypersetup{hidelinks}\textcolor{blue}{\cite{b22}}} that uses a simple, sequential architecture. It stacks multiple 3$\times$3 convolution layers with periodic max-pooling to reduce spatial size, followed by two fully-connected layers of 4096 neurons each and a final softmax classification layer{~\hypersetup{hidelinks}\textcolor{blue}{\cite{b22}}}. This "very deep" but straightforward design achieved high accuracy on ImageNet, but at the cost of a very large number of parameters. \textbf{ResNet} (Residual Network), introduced by He et al.{~\hypersetup{hidelinks}\textcolor{blue}{\cite{b23}}}, takes a different approach by using residual blocks with skip connections that alleviate vanishing gradients and allow training of much deeper networks{~\hypersetup{hidelinks}\textcolor{blue}{\cite{b23}}}. In a residual block, the input is added to the output of a few convolutional layers, enabling the network to learn residual functions rather than full transformations. This innovation allows ResNets to be extremely deep (50, 101 layers or more) without degradation in performance. For example, \textbf{ResNet-50} contains 50 weight layers and employs a "bottleneck" block design (stacking 3 convolutional layers per residual block instead of 2) to make the network both deeper and more efficient. Notably, despite its greater depth, ResNet-50 actually has far fewer parameters than VGG-19 because it replaces the large fully connected layers with a global average pooling and a single final linear layer. Both these CNN baselines were pre-trained on the ImageNet ILSVRC-2012 dataset (which has \textasciitilde1.2 million training images across 1000 classes){~\hypersetup{hidelinks}\textcolor{blue}{\cite{b22,b23}}}. We leverage the ImageNet-pretrained weights for each model and fine-tune them on our evaluation datasets, a transfer learning practice shown to greatly benefit performance by starting from learned general features. During fine-tuning, we experimented with different hyperparameter configurations, including learning rates of 0.0003 and 0.001, a weight decay of 0.01 and 0.001, batch sizes of 32 and 64, and data augmentations (e.g., random rotations, horizontal and vertical flips, and random resized cropping with scaling), to identify the optimal settings for each model and dataset.

\textbf{Transformer Baseline (ViT-Base)}. To represent transformer based vision models, we include the Vision Transformer \textbf{ViT-Base} as a baseline. ViT was first proposed by Dosovitskiy et al.{~\hypersetup{hidelinks}\textcolor{blue}{\cite{b13}}} as a "vanilla" transformer architecture for image recognition, analogous to the transformers used in NLP{~\hypersetup{hidelinks}\textcolor{blue}{\cite{b13}}}. Instead of convolutions, ViT processes an image by splitting it into a sequence of fixed-size patches (e.g. 16$\times$16 pixels), embedding each patch as a vector, and feeding those vectors into a standard Transformer encoder stack{~\hypersetup{hidelinks}\textcolor{blue}{\cite{b13}}}. The ViT-Base configuration (also known as ViT-B/16) consists of 12 transformer layers, with an embedding dimension of 768 and 12 self-attention heads in each layer. A learnable class token is prepended to the sequence of patch embeddings, and its output representation is used for classification. This design allows ViT to model long-range relationships between patches directly through self-attention, without any inductive bias of locality or translation invariance that CNNs have. ViT models have achieved competitive or state-of-the-art results on image classification, especially when trained on very large datasets{~\hypersetup{hidelinks}\textcolor{blue}{\cite{b24}}}. For our experiments, we use the ViT-Base model initialized with weights pre-trained on ImageNet (the same 1.2M images, 1000-class dataset as above) to ensure a fair comparison with CNN baselines. We then fine-tune the ViT on our tasks. This ImageNet-pretrained ViT baseline, often referred to as a "vanilla" ViT, provides a point of comparison to gauge the performance of a pure transformer architecture against CNN-based models using the same hyperparameters sweep. By evaluating VGG-19, ResNet-50, and ViT-Base side by side, we can assess how our method compares to both classical CNN architectures and modern transformer models, all under the same conditions.

\subsection{UltraSam Model}
UltraSam model{~\hypersetup{hidelinks}\textcolor{blue}{\cite{b1}}} uses the architecture of SAM{~\hypersetup{hidelinks}\textcolor{blue}{\cite{b76}}}, which leverages a ViT encoder to extract visual characteristics as tokens. A prompt encoder converts prompts (e.g., points or boxes) into object query tokens. These tokens engage with the visual feature tokens via a transformer decoder, facilitating reasoning and interaction between prompts and vision tokens. A mask head predicts many mask outputs, each associated with a projected Intersection over Union (IoU) score, facilitating the selection of the optimal expected mask. During a subsequent run through the decoder, the mask logits from the preceding iteration are encoded and added element-wise to the image embedding, therefore enhancing the mask prediction.

During pretraining, they emulated user prompts by randomly selecting either a point or a box with equal probability for each instance. The point was chosen randomly inside the instance mask, while the box is a perturbed version of the ground truth box annotation. To produce this noise, the two corners of the box were randomly shifted by up to 5 percent of the box's width and height.

We assess UltraSam's performance as the prior US foundation model using its feature extractor (ViT-encoder) as a pretrained backbone and attach a classification head for the downstream tasks (their reported results are presented in {\hypersetup{hidelinks}\textcolor{blue}{Table~\ref{tab:results_colored}}}).

\subsection{Evaluation Metrics}
To comprehensively assess the model's performance across all downstream tasks, three widely adopted quantitative metrics in medical image classification were used: precision, recall, and F1-score. These metrics provide a more nuanced evaluation than overall accuracy, because medical datasets typically suffer from class imbalance and different costs associated with false negatives and false positives. The metrics are defined as follows:

\begin{equation}
\text{Precision} = \frac{TP}{TP + FP}
\label{eq:precision}
\end{equation}
\begin{equation}
\text{Recall} = \frac{TP}{TP + FN}
\label{eq:recall}
\end{equation}
\begin{equation}
\text{F1-score} = \frac{2 \times \text{Precision} \times \text{Recall}}{\text{Precision} + \text{Recall}}
\label{eq:f1-score}
\end{equation}

where TP, FP, and FN are the number of true positive, false positive, and false negative predictions, respectively.

\textbf{Precision} ({\hypersetup{hidelinks}\textcolor{blue}{Eq.~\ref{eq:precision}}}) quantifies the model’s ability to avoid false positives, that is, the proportion of correctly predicted positive samples among all predicted positives. This metric becomes clinically relevant in cases where a false diagnosis may lead to unnecessary follow-up procedures or patients' anxiety.

\textbf{Recall (Sensitivity)} ({\hypersetup{hidelinks}\textcolor{blue}{Eq.~\ref{eq:recall}}}) is the proportion of positive cases the model correctly identifies. In diagnostic applications, recall is particularly important because missing a true abnormality may have significant clinical implications. 

The \textbf{F1-score} ({\hypersetup{hidelinks}\textcolor{blue}{Eq.~\ref{eq:f1-score}}}) is a harmonic mean of precision and recall, balancing their trade-off. This single scalar value is particularly useful when dataset classes are imbalanced, which is typical in US datasets where pathological cases may be underrepresented.

All reported metrics were computed on the test sets defined by each dataset’s official split, and performance comparisons between baselines, UltraSam, and USF-MAE were made using identical evaluation protocols.

\begin{table*}[htbp]
\centering
\caption{Classification performance on BUS-BRA, MMOTU-2D, and GIST514-DB datasets. 
Best results in \textcolor{green}{green}, second best in \textcolor{orange}{orange}, and worst in \textcolor{red}{red}.}
\label{tab:results_colored}
\setlength{\tabcolsep}{10pt} % Increase column spacing if needed
\renewcommand{\arraystretch}{1.3} % Increase row height for readability
\begin{tabular}{lccc ccc ccc}
\toprule
\multirow{2}{*}{\textbf{Model}} &
\multicolumn{3}{c}{\textbf{BUS-BRA}} &
\multicolumn{3}{c}{\textbf{MMOTU-2D}} &
\multicolumn{3}{c}{\textbf{GIST514-DB}} \\
\cmidrule(lr){2-4} \cmidrule(lr){5-7} \cmidrule(lr){8-10}
& Prec. & Rec. & F1-score & Prec. & Rec. & F1-score & Prec. & Rec. & F1-score \\
\midrule
VGG-19 & \cellcolor{red!50}60.8 & \cellcolor{red!50}50.8 & \cellcolor{red!50}55.4
       & 62.7 & 62.7 & 62.7
       & \cellcolor{red!50}69.6 & 88.1 & 77.3 \\
ResNet-50 & 69.1 & 77.0 & 72.9 & 73.2 & 73.2 & 73.2 & 74.0 & \cellcolor{orange!60}90.8 & \cellcolor{orange!60}81.3 \\
ViT-Base & 71.4 & 65.6 & 68.4 & \cellcolor{orange!60}74.7 & \cellcolor{orange!60}74.7 & \cellcolor{orange!60}74.7 & 73.7 & 89.3 & 80.6 \\
UltraSam{~\hypersetup{hidelinks}\textcolor{blue}{\cite{b1}}} 
  & \cellcolor{green!40}87.6 & \cellcolor{green!40}87.5 & \cellcolor{green!40}87.5
  & \cellcolor{red!50}62.6 & \cellcolor{red!50}62.4 & \cellcolor{red!50}62.0
  & \cellcolor{green!40}78.2 & \cellcolor{red!50}74.2 & \cellcolor{red!50}73.5 \\
\textbf{USF-MAE (ours)} 
  & \cellcolor{orange!60}81.3 & \cellcolor{orange!60}82.0 & \cellcolor{orange!60}81.6
  & \cellcolor{green!40}79.6 & \cellcolor{green!40}79.6 & \cellcolor{green!40}79.6
  & \cellcolor{orange!60}75.1 & \cellcolor{green!40}91.3 & \cellcolor{green!40}82.4 \\
\bottomrule
\end{tabular}
\end{table*}

\section{Results}
{\hypersetup{hidelinks}\textcolor{blue}{Table~\ref{tab:results_colored}}} summarizes the quantitative performance of USF-MAE versus two standard architectures (VGG-19 and ResNet-50), ViT-Base, the UltraSam foundation model{~\hypersetup{hidelinks}\textcolor{blue}{\cite{b1}}}, and our USF-MAE foundation model, on three representative US classification tasks: 1) BUS-BRA (breast cancer){~\hypersetup{hidelinks}\textcolor{blue}{\cite{b15}}}, 2) MMOTU-2D (ovarian tumor subtype){~\hypersetup{hidelinks}\textcolor{blue}{\cite{b17}}}, and 3) GIST514-DB (gastrointestinal stromal tumor){~\hypersetup{hidelinks}\textcolor{blue}{\cite{b16}}}.

% On the BUS-BRA dataset , USF-MAE reached a precision of 81.3\%, a recall of 82.0\%, and an F1-score of 81.6\%, outperforming all conventional baselines and approaching the performance of the prior US foundation model UltraSam (F1-score of 87.5\%). USF-MAE was the highest-performing model on the MMOTU-2D dataset, surpassing the second highest-performing model (ViT-Base) by nearly 5\% in precision, recall, and F1-score. On the GIST514-DB dataset, USF-MAE was also the best-performing model and achieved the highest recall and F1-score (91.3\% and 82.4\%, respectivully). UltraSam had the best precision (78.2\%), but was the worst performaing model on the remaining metrics. While UltraSam performed well on the BUS-BRA dataset, it was the worst performing model for the MMOTU-2D and GIST514-DB datasets and we will get into further details in the Disucssion section.

\begin{figure*}
	\centering
	\includegraphics[width=.8\textwidth]{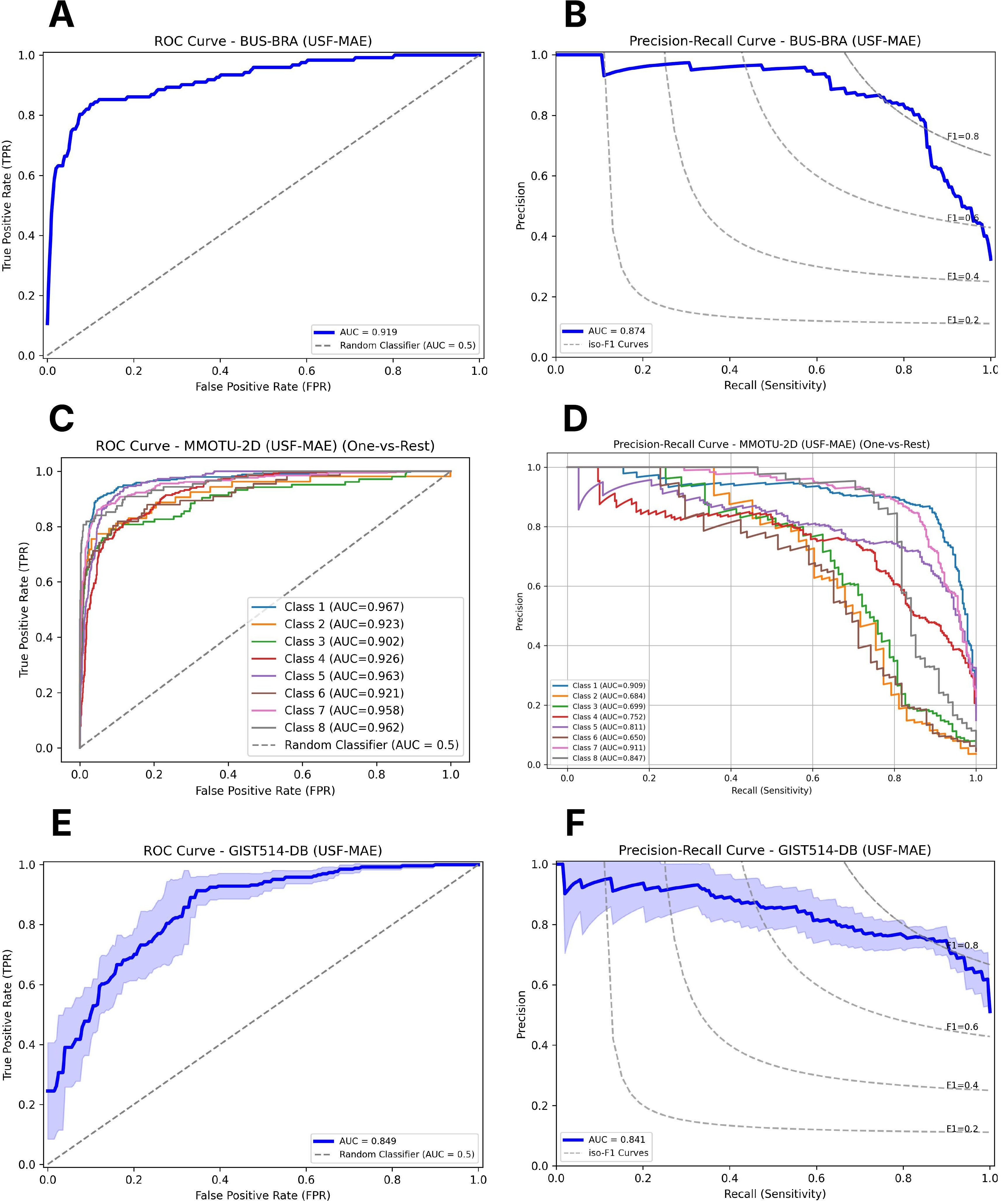}
	\caption{ROC and Precision-Recall curves of the USF-MAE model evaluated on three datasets. (A-B) Binary classification performance on BUS-BRA. (C-D) One-vs-Rest multi-class performance on MMOTU-2D. (E-F) Binary classification performance on GIST514-DB. AUC values are reported for each curve.}
	\label{FIG:curves_figure}
\end{figure*}

On the BUS-BRA dataset, USF-MAE achieved a precision of 81.3\%, recall of 82.0\%, and F1-score of 81.6\%, outperforming all conventional baselines. USF-MAE attained the highest overall performance on the MMOTU-2D dataset, exceeding the second-best model (ViT-Base) by nearly 5\% across all three evaluation metrics. On the GIST514-DB dataset, USF-MAE achieved the best overall results, recording the highest recall (91.3\%) and F1-score (82.4\%). Although UltraSam achieved the highest precision (78.2\%) on GIST514-DB, it performed substantially worse on recall and F1-score, ranking as the lowest among all models on those metrics. To further characterize the performance of USF-MAE beyond the scalar metrics reported in {~\hypersetup{hidelinks}\textcolor{blue}{Table.~\ref{tab:results_colored}}}, we analyzed its Receiver Operating Characteristic (ROC) and Precision-Recall (PR) curves across the three downstream datasets. As shown in{~\hypersetup{hidelinks}\textcolor{blue}{Fig.~\ref{FIG:curves_figure}}}, USF-MAE maintains stable discriminative behaviour across both binary and multi-class settings, with high AUC values observed for BUS-BRA, MMOTU-2D, and GIST514-DB. These curves illustrate the confidence-recall trade-offs of the model and provide a more detailed view of its decision boundary behaviour than single-value metrics alone.

\section{Discussion}
\subsection{Interpretation of Results}
Results demonstrate that USF-MAE serves as an effective foundation model for efficient, domain-adaptive labeling in US imaging. They illustrate the USF-MAE model’s ability to extract high-quality, transferable representations across various anatomical regions and image acquisition conditions, as evidenced by consistent improvements across three downstream tasks: 1) BUS-BRA, 2) MMOTU-2D, and 3) GIST514-DB.

A key distinction between the foundation models USF-MAE and UltraSam lies in their training paradigm. UltraSam was trained in a supervised manner using labeled US images that included pathology annotations (e.g., breast and ovarian tumor labels). In contrast, USF-MAE is entirely self-supervised, learning from unlabeled US scans by reconstructing masked image patches.  This approach enables the use of large amounts of unannotated clinical US data that are typically excluded from traditional training workflows. Such data are abundant yet underutilized in healthcare institutions, as labeling requires domain expertise and significant time and money investments. Leveraging unlabeled data provides a scalable and cost-efficient way to improve model performance, which is especially valuable in medical imaging where labeled data are often limited but unlabeled data are abundant. For example, The Ottawa Hospital (TOH) and similar tertiary centers maintain large clinical repositories of routinely collected US images, including thousands of obstetrical scans that include images of the fetal anatomy, placenta and uterus. These routine US images are overwhelmingly from normal uncomplicated pregnancies, but also include cases of fetal congenital anomalies (e.g., congenital heart defects) and other important adverse outcomes such as preeclampsia (PE).  These images are mostly unlabelled, or labelling is not directly attached to US images (e.g., recorded in clinical narrative notes, ICD codes in EMR databases). These images typically remain unlabeled due to annotation costs, workload, and privacy considerations, unless they are directly used in institutional research ethics board-approved research. These large clinical repositories represent an untapped resource for further self-supervised pre-training of foundation models such as the USF-MAE checkpoint, which can be safely conducted behind hospital firewalls without exposing/disclosing personal health information. Thus, USF-MAE demonstrates strong downstream performance even when fine-tuned with a relatively small labeled dataset, showing high label efficiency and positioning it as a scalable foundation model across domains where labeled US data are scarce but unlabeled data are abundant.
 
Generally, these results show that self-supervised pretraining on big, unlabeled US data gives strong, useful representations that match the performance of fully supervised foundation models trained with labeled pathology data. USF-MAE achieves this while keeping good generalization across different anatomical areas and diagnostic tasks, highlighting its possible use as a scalable and label-efficient foundation model for US imaging.

\textbf{1) Comparative Analysis of Model Performance:}
The performance patterns observed across the three evaluation tasks further illuminate the differences between self-supervised and supervised foundation models. USF-MAE substantially outperformed the conventional CNN-based architectures (VGG-19, ResNet-50), the vanilla ViT-Base baseline, and the UltraSam foundation model while UltraSam outperformed the other models only in the BUS-BRA dataset. Our model improvement can be attributed to being pre-trained on large-scale US images and the ability of foundation-style encoders to capture global contextual representations and long-range dependencies within US images, rather than relying solely on localized textural cues. The masked reconstruction objective of USF-MAE, in particular, enables learning of more noise-tolerant structural features, which are essential in US, where speckle artifacts and variable probe orientations often obscure fine details.

However, an interesting observation arises when comparing the performance of USF-MAE and UltraSam. UltraSam achieved the highest scores on the BUS-BRA breast cancer dataset (precision = 87.6\%, recall = 87.5\%, F1 = 87.5\%), while performing substantially worse on the MMOTU-2D and GIST514-DB classification tasks (F1 = 62.0\% and 73.5\%, respectifully). This discrepancy can be explained by the nature of UltraSam’s supervised pretraining, which included multiple breast US datasets in its labeled training corpus. As a result, the model effectively "saw" breast lesion patterns during pretraining and thus benefited from prior task familiarity when fine-tuned on BUS-BRA. In contrast, ovarian and gastrointestinal stromal tumours US data were either minimally represented or absent from UltraSam’s training set, limiting its ability to generalize to the MMOTU-2D and GIST514-DB datasets. In addition, it is worth noting that the UltraSam study also reported ResNet-50 results, which were lower than those obtained in our experiments. This difference may stem from variations in fine-tuning configurations, such as learning rate, batch size, or other hyperparameter settings and data augmentation techniques, as well as potential differences in preprocessing or data sampling strategies.

By comparison, USF-MAE was trained using a self-supervised objective on unlabeled data that encompassed a broader range of anatomical domains. Despite never being explicitly trained to recognize breast lesions, it achieved performance very close to UltraSam on BUS-BRA and substantially higher on MMOTU-2D and GIST514-DB. This suggests that the representations learned by USF-MAE are more domain-agnostic and generalizable, as they are driven by intrinsic US statistics rather than class-specific supervision. Consequently, while supervised foundation models such as UltraSam can achieve strong task-specific performance when label distributions align, they are prone to domain overfitting, leading to weaker generalization in unseen anatomical contexts.

The superior overall performance of USF-MAE across all three tasks, therefore, highlights the strength of self-supervised, modality-focused pretraining in capturing the shared structural priors of US data. By emphasizing reconstruction based representation learning rather than supervised discrimination, USF-MAE develops a more versatile feature space that can be readily adapted to new clinical domains without requiring retraining on large labeled datasets.

\textbf{2) Why Self-Supervised Pre-Training Benefits US:} US is a difficult imaging modality because it has a lot of speckle noise, a restricted field of view, and is dependent on the operator. All of these things make it harder to use models that were trained on general images. Traditional CNN and Transformer architectures pre-trained on ImageNet learn texture and edge statistics typical of photographic images, which are fundamentally different from the acoustic scattering patterns in US. Consequently, transferring such models to sonographic data often yields suboptimal representations.

By contrast, USF-MAE learns directly from the US domain without relying on manual labels. Through masked image reconstruction, the model learns to infer missing structural information by modeling the statistical dependencies between neighbouring patches. This process encourages the encoder to capture both global contextual structures (organ geometry, echotexture gradients) and local fine-grained features (lesion boundaries, tissue speckle distributions), effectively learning noise-resilient representations. Self-supervised learning forces the model to learn the inherent statistics of the modality itself, rather than supervised pretraining which encourages a bias toward discriminative features relevant to specific labeled classes. This property makes USF-MAE particularly robust to domain shifts such as differences in probe type, acquisition angle, or patient anatomy.

\textbf{3) Impact of Large-Scale Pretraining on Downstream Tasks:}
Though USF-MAE does not use labels in its pretraining, seeing hundreds of thousands of US scans from muscles, tissues, livers, and different organs gives great representational diversity. Each scan inherently encompasses data regarding organ morphology, echogenicity patterns, and acoustic artifacts, enabling the model to acquire modality-specific priors advantageous for numerous clinical applications. The results in{~\hypersetup{hidelinks}\textcolor{blue}{Table~\ref{tab:results_colored}}} show that this large-scale pretraining leads to better performance down the line, even with only minimal fine-tuning. This phenomenon corresponds with the extensive literature on self-supervised representation learning, indicating that increasing the pretraining corpus, as opposed to the label set, enhances generalization and accelerates convergence during fine-tuning.

Furthermore, the ability to pretrain on unlabeled US data enables continuous model improvement as new scans become available. Hospitals routinely acquire terabytes of US data annually, the vast majority of which remain unlabeled. The self-supervised objective used in USF-MAE enables seamless integration of such data into future pretraining rounds without the need for manual annotation. It makes USF-MAE naturally extensible: new anatomical regions, patient populations, or imaging protocols added to the model will dynamically change its representation space.

\textbf{4) Transferability Across Anatomically Diverse Tasks:}
The outstanding performance of USF-MAE on the three evaluation datasets strongly suggests that it can be used across different types of anatomy. Even though they were trained without labels, the representations learned during pretraining worked well for a wide range of tasks, such as classifying lesions in breast and gastrointestinal tissues. These tasks differ significantly in visual appearance, spatial scale, and semantic granularity; nevertheless, USF-MAE consistently outperformed both CNN and ViT baselines.

This generalization capacity arises from the transformer’s global receptive field and the self-supervised reconstruction objective, which together encourage modeling of long-range spatial dependencies rather than localized features alone. Because of this, USF-MAE can use contextual cues from surrounding anatomy to better tell the difference between subtle structural patterns, like the edges of small lesions or the relative orientation of heart chambers.

In practical terms, this means that USF-MAE can be used as a universal feature extractor for US, which means that it can be used in many different applications with little or no changes needed for each one. The pretrained encoder can be fine-tuned quickly even on small, specialized datasets, whether it is used for disease detection, organ segmentation, or view classification. This is especially useful in medical settings where there is a lack of labeled data and a lot of class imbalance.

\textbf{5) Broader Implications:}
These findings demonstrate that \textbf{USF-MAE addresses a significant deficiency in modern medical AI pipelines}: the capacity to learn from unlabeled, modality-specific data at scale. Its success shows that self-supervised transformer frameworks can achieve representation quality similar to fully supervised models when used directly in the US field. They are also flexible, data-efficient, and scalable. This not only reduces the dependency on large annotated corpora but also opens a pathway toward continual learning, where foundation models can evolve alongside the clinical data streams generated by hospitals worldwide.

\subsection{Clinical Significance}
The advancement of USF-MAE has significant clinical potential for the future of AI in US imaging. USF-MAE solves one of the biggest problems in medical AI: the lack of high-quality labeled datasets that are both cheap and easy to find. It does this by making it possible to learn effective representations from unlabeled data. Labeling US images is especially hard because it requires skilled sonographers or radiologists, takes a lot of time, and is very subjective. Because of this, even major hospitals and clinics have huge collections of unlabeled US scans that are not being used for developing AI models.

USF-MAE learns domain-relevant features directly from unlabeled data through self-supervised pretraining. This greatly lessens the need for hand-labelled datasets. When fine-tuned with a small set of data specific to the task, the pre-trained encoder can well change these broad representations into results that match or even beat those of fully guided models trained on large annotated datasets. This property makes USF-MAE an ideal choice for efficiently annotating US.

From the clinical perspective, the transform-based architecture allows modeling global contextual relationships within US scans, thereby achieving improved sensitivity to subtle shape variations, which are essentially required in diverse diagnostic scenarios. For instance, USF-MAE could be modified to detect PE by analyzing uteroplacental flow patterns or placental morphology{~\hypersetup{hidelinks}\textcolor{blue}{\cite{b77}}}, identify cystic hygroma in early fetal imaging through the recognition of characteristic nuchal fluid collections{~\hypersetup{hidelinks}\textcolor{blue}{\cite{b78,b79}}}, or classify congenital kidney malformations such as hydronephrosis and multicystic dysplastic kidneys{~\hypersetup{hidelinks}\textcolor{blue}{\cite{b80,b81}}} by fine-tuning the pretrained checkpoint on small, labeled subsets. Its ability to model fine structural detail makes it a good tool for finding fetal brain anomalies, since early detection of ventriculomegaly or cortical malformations{~\hypersetup{hidelinks}\textcolor{blue}{\cite{b82}}} can help doctors decide what to do next.

Because self-supervised pretraining can be used on a large scale, USF-MAE can also be improved over time as more US data becomes available from different clinical departments. This makes it possible for hospitals to adapt their models to their own needs and share pretraining data without sharing patient-sensitive data.

The clinical importance of USF-MAE resides in its capacity to democratize the development of US AI. By separating model performance from the availability of labeled data, it allows smaller institutions, research groups, and developing regions to create reliable diagnostic models that are specific to their patient populations. In this way, USF-MAE is a step toward making US AI accessible, scalable, and adaptable to clinical settings, which could help improve point-of-care diagnostics and personalized patient care.

\subsection{Future Directions}
The results gained from USF-MAE are highly promising, but there are other avenues for further development as well as validation of the model in broader clinical environments. One important goal is to add more and bigger US data from different institutions and imaging systems to the pretraining corpus. This study used a large collection of open-source datasets, but the generalization of any foundation model depends on how diverse its training data is. Added US collections across various clinical settings will teach the model differences in acquisition protocols, transducer types, and patient populations. This will make it more robust across institutions.

In this context, TOH maintains a large collection of unlabeled echocardiography and obstetrical US scans, including thousands of studies completed as part of clinical care. If the model were to be further pre-trained on these scans, it would improve at understanding anatomy, including how the heart moves and how the placenta typically looks. This would improve utility in specialized diagnostic tasks like classifying congenital heart disease (CHD), assessing the risk of PE, identifying cystic hygromas, and identifying kidney anomalies. By utilizing extensive amounts of unlabeled US data from these and various other fields, USF-MAE can gradually develop into a more inclusive representation model that encompasses the entirety of clinical sonography.

Another important area of focus is clinical integration and validation from other sources. The pretrained model will undergo prospective evaluation within ongoing research ethics board-research initiatives at the Ottawa Hospital Research Institute (OHRI). In prospective studies, we will validate the model’s ability to detect structural anomalies, including cystic hygroma and renal anomalies, identify placental and fetal features associated with PE, and aid in early classification of congenital heart disease based on fetal echocardiogram images and cineloops. These studies will help ascertain the diagnostic reproducibility and reliability of this model in the real clinical workflow, as factors such as patient heterogeneity, device differences, and image variability often pose great challenges.

Our long-term plan is to gradually evolve USF-MAE into an adaptive foundation US model by means of a staged pretraining, external validation, and clinical translation process that would later be utilized as a consistent framework for interpretation of obstetrical US images, where the confidence level over diagnosis varies greatly between different applications.

\section{Conclusion}
In this study, we introduced USF-MAE, the \textit{Ultrasound Self-Supervised Foundation Model with Masked Autoencoding}, as the first large-scale, self-supervised transformer framework trained exclusively on US data. Leveraging $\sim$370,000 images from 46 open-source datasets (OpenUS-46), USF-MAE demonstrated the ability to learn rich, domain-specific representations directly from unlabeled US scans. When fine-tuned on three independent downstream datasets, BUS-BRA, MMOTU-2D, and GIST514-DB, the model consistently outperformed conventional convolutional and transformer baselines, achieving strong cross-anatomical generalization despite never being trained with labels.

Compared to the supervised foundation model UltraSam{~\hypersetup{hidelinks}\textcolor{blue}{\cite{b1}}}, USF-MAE achieved superior performance while relying solely on self-supervised pretraining, underscoring its efficiency in leveraging unlabeled data. These findings highlight the potential of self-supervised masked autoencoding to overcome one of the primary barriers in US machine learning: the scarcity of annotated data.

Beyond its current results, the USF-MAE framework provides a scalable foundation for continual learning, capable of integrating new, unlabeled US data from both public and institutional sources without requiring manual annotation. This adaptability positions USF-MAE as a sustainable and extensible platform for developing robust, label-efficient US models across diverse diagnostic domains. All USF-MAE pretrained weights, the complete list of OpenUS-46 dataset links, and the pretraining code are publicly released in our \href{https://github.com/Yusufii9/USF-MAE}{\textcolor{blue}{\textbf{GitHub Repository}}}\\(https://github.com/Yusufii9/USF-MAE) to further research collaboration in the US AI community.

\section*{Declaration of competing interest}
The authors declare that they have no known competing financial interests or personal relationships that could have appeared to influence the work reported in this paper.

%\vskip3pt


\begin{thebibliography}{0}
% \bibitem{b0} J. Kessmann \textit{et al.}, "Hirschsprung disease: diagnosis and management," \textit{American Family Physician}, vol. 74, no. 8, pp. 1319-1322, 2006.

% \bibitem{b1} Meyer A, Murali A, Zarin F, Mutter D, Padoy N. Ultrasam: a foundation model for ultrasound using large open-access segmentation datasets. \textit{Int J Comput Assist Radiol Surg}. 2025 Sep 11. doi: 10.1007/s11548-025-03517-8. Epub ahead of print. PMID: 40932576.

\bibitem{b1} A. Meyer, A. Murali, F. Zarin, D. Mutter, and N. Padoy, “UltraSam: A foundation model for ultrasound using large open-access segmentation datasets,” \textit{Int. J. Comput. Assist. Radiol. Surg.}, Sep. 11, 2025, doi: 10.1007/s11548-025-03517-8.

\bibitem{b2} NHS England, “Diagnostic Imaging Dataset Statistical Release: Provisional monthly statistics, February 2016 to February 2017,” Version 1.0, \textit{Operational Information for Commissioning, NHS England}, Leeds, UK, 13 Jul. 2017.

\bibitem{b3} D. Won, J. Walker, R. Horowitz, S. Bharadwaj, E. Carlton, and H. Gabriel, “Sound the alarm: The sonographer shortage is echoing across healthcare,” \textit{J. Ultrasound Med.}, vol. 43, no. 7, pp. 1289–1301, Jul. 2024, doi: 10.1002/jum.16453.

\bibitem{b4} Q. Kang, Q. Lao, J. Gao, W. Bao, Z. He, C. Du, Q. Lu, and K. Li, “URFM: A general ultrasound representation foundation model for advancing ultrasound image diagnosis,” \textit{iScience}, vol. 28, no. 8, p. 112917, Aug. 2025, doi: 10.1016/j.isci.2025.112917.

\bibitem{b5} J. K. Hoang, J. Middleton, L. Farjat, A. R. Langer, S. H. Wu, and N. R. Beland, “Interobserver variability of sonographic features used in the American College of Radiology Thyroid Imaging Reporting and Data System,” \textit{AJR Am. J. Roentgenol.}, vol. 211, no. 1, pp. 162–167, Jul. 2018, doi: 10.2214/AJR.17.19192.

\bibitem{b6} S. Strauss, Z. Gavish, E. Gottlieb, and M. Katsnelson, “Interobserver and intraobserver variability in the sonographic assessment of the presence and severity of fatty liver,” \textit{J. Clin. Ultrasound}, vol. 35, no. 3, pp. 182–187, Mar.–Apr. 2007, doi: 10.2214/AJR.07.2123.

\bibitem{b7} K. He, C. Gan, Z. Li, I. Rekik, Z. Yin, W. Ji, Y. Gao, Q. Wang, J. Zhang, and D. Shen, “Transformers in medical image analysis,” \textit{Intelligent Medicine}, vol. 3, no. 1, pp. 59–78, 2023, doi: 10.1016/j.imed.2022.07.002.

\bibitem{b8} K. Song, J. Feng, and D. Chen, “A survey on deep learning in medical ultrasound imaging,” \textit{Front. Phys.}, vol. 12, 2024, doi: 10.3389/fphy.2024.1398393.

\bibitem{b9} J. Deng, W. Dong, R. Socher, L. -J. Li, Kai Li and Li Fei-Fei, "ImageNet: A large-scale hierarchical image database," \textit{2009 IEEE Conference on Computer Vision and Pattern Recognition}, Miami, FL, USA, 2009, pp. 248-255, doi: 10.1109/CVPR.2009.5206848.

\bibitem{b10} X. Mei, Z. Liu, P. M. Robson, B. Marinelli, M. Huang, A. Doshi, A. Jacobi, C. Cao, K. E. Link, T. Yang, Y. Wang, H. Greenspan, T. Deyer, Z. A. Fayad, and Y. Yang, “RadImageNet: An open radiologic deep learning research dataset for effective transfer learning,” \textit{Radiol. Artif. Intell.}, vol. 4, no. 5, Sep. 2022, Art. no. e220058, doi: 10.1148/ryai.210315.

\bibitem{b11} Y. Chen, C. Zhang, L. Liu, C. Feng, C. Dong, Y. Luo, and X. Wan, “USCL: Pretraining deep ultrasound image diagnosis model through video contrastive representation learning,” \textit{in Proc. Med. Image Comput. Comput. Assist. Interv. (MICCAI)}, vol. 12908, Cham, Switzerland: Springer, 2021, pp. 638–648, doi: 10.1007/978-3-030-87237-3\_60. 

\bibitem{b12} H. E. Kim, A. Cosa-Linan, N. Santhanam, M. Jannesari, M. E. Maros, and T. Ganslandt, “Transfer learning for medical image classification: A literature review,” \textit{BMC Med. Imaging}, vol. 22, no. 69, pp. 1–13, May 2022, doi: 10.1186/s12880-022-00793-7. 

\bibitem{b13} A. Dosovitskiy, L. Beyer, A. Kolesnikov, D. Weissenborn, X. Zhai, T. Unterthiner, M. Dehghani, M. Minderer, G. Heigold, S. Gelly, J. Uszkoreit, and N. Houlsby, “An image is worth 16×16 words: Transformers for image recognition at scale,” \textit{in Proc. Int. Conf. Learn. Representations (ICLR)}, 2021. Available: https://openreview.net/forum?id=YicbFdNTTy. 

\bibitem{b14} J. G. de Almeida, L. C. Alberich, G. Tsakou, K. Marias, M. Tsiknakis, K. Lekadir, L. Marti-Bonmati, and N. Papanikolaou, “Foundation models for radiology—the position of the AI for Health Imaging (AI4HI) network,” \textit{Insights Imaging}, vol. 16, no. 1, p. 168, Aug. 2025, doi: 10.1186/s13244-025-02056-9.

\bibitem{b15} W. Gómez-Flores, M. J. Gregorio-Calas, and W. C. de Albuquerque Pereira, “BUS-BRA: A breast ultrasound dataset for assessing computer-aided diagnosis systems,” \textit{Med. Phys.}, vol. 51, no. 4, pp. 3110–3123, Apr. 2024, doi: 10.1002/mp.16812.

\bibitem{b16} Q. He, S. Bano, J. Liu, W. Liu, D. Stoyanov, and S. Zuo, "Query2: Query over queries for improving gastrointestinal stromal tumour detection in an endoscopic ultrasound," \textit{Computers in Biology and Medicine}, vol. 152, p. 106424, 2023.

\bibitem{b17} Q. Zhao, S. Lyu, W. Bai, L. Cai, B. Liu, G. Cheng, M. Wu, X. Sang, M. Yang, and L. Chen, “MMOTU: A multi-modality ovarian tumor ultrasound image dataset for unsupervised cross-domain semantic segmentation,” \textit{arXiv preprint} arXiv:2207.06799, 2022.

% \bibitem{b18} C. Stoean, R. Stoean, M. Hotoleanu, D. Iliescu, C. Patru, and R. Nagy, “An assessment of the usefulness of image pre-processing for the classification of first trimester fetal heart ultrasound using convolutional neural networks,” \textit{in Proc. 25th Int. Conf. Syst. Theory, Control Comput. (ICSTCC)}, Iasi, Romania, 2021, pp. 242–248, doi: 10.1109/ICSTCC52150.2021.9606852.

\bibitem{b19} Y. Megahed, . A. . Fuller, . S. . Abou-Alwan, . D. El Demellawy, and A. Chan, “Segmentation of Muscularis Propria in Colon Histopathology Images Using Vision Transformers for Hirschsprung’s Disease”, \textit{CMBES Proc.}, vol. 47, no. 1, May 2025.

\bibitem{b20} Y. Amethiya, P. Pipariya, S. Patel, and M. Shah, “Comparative analysis of breast cancer detection using machine learning and biosensors,” \textit{Intelligent Medicine}, vol. 2, no. 2, pp. 69–81, 2022, doi: 10.1016/j.imed.2021.08.004.

\bibitem{b21} Y. Xie, L. Gu, T. Harada, J. Zhang, Y. Xia, and Q. Wu, "Rethinking masked image modelling for medical image representation," \textit{Medical Image Analysis}, vol. 98, p. 103304, 2024. doi: 10.1016/j.media.2024.103304.

\bibitem{b22} K. Simonyan and A. Zisserman, “Very deep convolutional networks for large-scale image recognition,” \textit{arXiv preprint} arXiv:1409.1556, 2015. Available: https://arxiv.org/abs/1409.1556.

\bibitem{b23} K. He, X. Zhang, S. Ren, and J. Sun, “Deep residual learning for image recognition,” in Proc. \textit{IEEE Conf. on Computer Vision and Pattern Recognition (CVPR)}, 2016.

\bibitem{b24} L. Scabini, A. Sacilotti, K. M. Zielinski, L. C. Ribas, B. De Baets, and O. M. Bruno, “A comparative survey of vision transformers for feature extraction in texture analysis,” \textit{arXiv preprint} arXiv:2406.06136v1, 2024. Available: https://arxiv.org/abs/2406.06136.

\bibitem{b25} S. Singh, P. K. Jain, N. Sharma, M. Pohit, and S. Roy, “Atherosclerotic plaque classification in carotid ultrasound images using machine learning and explainable deep learning,” \textit{Intelligent Medicine}, vol. 4, no. 2, pp. 83–95, 2024, doi: 10.1016/j.imed.2023.05.003.

\bibitem{b26} S.-Q. Wang, C.-N. Jiao, Y.-L. Gao, X.-C. Cui, Y.-L. Wang, and J.-X. Liu, “Deep association analysis framework with multi-modal attention fusion for brain imaging genetics,” \textit{Medical Image Analysis}, vol. 107, p. 103827, 2026, doi: 10.1016/j.media.2025.103827.

\bibitem{b27} X. Zhang, J. Feng, P. Liu, M. Han, Y. Kang, J. Zhu, L. Wang, X. Wang, S. Ali, and L. Zhang, “Nested resolution mesh-graph CNN for automated extraction of liver surface anatomical landmarks,” \textit{Medical Image Analysis}, vol. 107, p. 103825, 2026, doi: 10.1016/j.media.2025.103825.


% datasets references
\bibitem{b28} X. P. Burgos-Artizzu, "FETAL\_PLANES\_DB: Common maternal-fetal ultrasound images", \textit{Nature Scientific Reports}, vol. 10. Zenodo, p. 10200, Jun. 23, 2020. doi: 10.5281/zenodo.3904280.

\bibitem{b29} J. Bai, "PSFHS", PSFHS: Intrapartum ultrasound image dataset for AI-based segmentation of pubic symphysis and fetal head. \textit{Zenodo}, Apr. 13, 2024. doi: 10.5281/zenodo.10969427.

\bibitem{b30} H. Elmekki, A. Alagha, H. Sami, A. Spilkin, A. M. Zanuttini, E. Zakeri, J. Bentahar, L. Kadem, W.-F. Xie, P. Pibarot, R. Mizouni, H. Otrok, S. Singh, and A. Mourad, "CACTUS: An open dataset and framework for automated cardiac assessment and classification of ultrasound images using deep transfer learning," \textit{Computers in Biology and Medicine}, vol. 190, p. 110003, 2025. doi: 10.1016/j.compbiomed.2025.110003.

\bibitem{b31} H. Gong, J. Chen, G. Chen, H. Li, G. Li, and F. Chen, "Thyroid region prior guided attention for ultrasound segmentation of thyroid nodules," \textit{Computers in Biology and Medicine}, vol. 155, p. 106389, 2023. doi: 10.1016/j.compbiomed.2022.106389.

\bibitem{b32} Y. Lu, J. Bai, M. Zhou, and M. Zhou, "JNU-IFM," \textit{figshare, Dataset}, 2021. doi: 10.6084/m9.figshare.14371652.v2.

\bibitem{b33} A. Hann, L. Bettac, M. M. Haenle, T. Graeter, A. W. Berger, J. Dreyhaupt, D. Schmalstieg, W. G. Zoller, and J. Egger, "Algorithm guided outlining of 105 pancreatic cancer liver metastases in ultrasound," \textit{Scientific Reports}, vol. 7, p. 12779, 2017. doi: 10.1038/s41598-017-12925-z.

\bibitem{b34} E. Iarussi, A. Díaz, and I. Larrabide, "US Simulation \& Segmentation," \textit{Kaggle}. Available: https://www.kaggle.com/datasets/ignaciorlando/ussimandsegm.

\bibitem{b35} M. S. Sappia, "ACOUSLIC-AI : Abdominal Circumference Operator-agnostic UltraSound measurement in Low-Income Countries using Artificial Intelligence". \textit{Zenodo}, 2024. doi: 10.5281/zenodo.12697994.

\bibitem{b36} Xu Yiming, "Annotated Ultrasound Liver images". \textit{Zenodo}, 2022. doi: 10.5281/zenodo.7272660.

\bibitem{b37} N. Wiedemann, D. de Korte-de Boer, M. Richter, S. van de Weijer, C. Buhre, F. A. M. Eggert, S. Aarnoudse, L. Grevendonk, S. Röber, C. M. E. Remie, W. Buhre, R. Henry, and J. Born, "COVID-BLUeS – A prospective study on the value of AI in lung ultrasound analysis," \textit{IEEE Journal of Biomedical and Health Informatics}, vol. 29, no. 9, pp. 6301–6310, 2025. doi: 10.1109/JBHI.2025.3543686.

\bibitem{b38} A. Tyagi, A. Tyagi, M. Kaur, R. Aggarwal, K. D. Soni, J. Sivaswamy, and A. Trikha, "Nerve Block Target Localization and Needle Guidance for Autonomous Robotic Ultrasound Guided Regional Anesthesia," \textit{in 2024 IEEE/RSJ International Conference on Intelligent Robots and Systems (IROS)}, pp. 5867–5872, 2024. doi: 10.1109/IROS58592.2024.10801467.

\bibitem{b39} A. Montoya, Hasnin, kaggle446, Shirzad, W. Cukierski, and yffud, "Ultrasound Nerve Segmentation," \textit{Kaggle}, 2016. Available: https://kaggle.com/competitions/ultrasound-nerve-segmentation.

\bibitem{b40} A. Pawłowska, A. Ćwierz-Pieńkowska, A. Domalik, D. Jaguś, P. Kasprzak, R. Matkowski, Ł. Fura, A. Nowicki, and N. Zolek, "A Curated Benchmark Dataset for Ultrasound Based Breast Lesion Analysis (Breast-Lesions-USG), version 1," \textit{The Cancer Imaging Archive}, 2024. doi: 10.7937/9WKK-Q141.

\bibitem{b41} A. Iqbal, "BUS\_UC," \textit{Mendeley Data}, version 1, 2023. doi: 10.17632/3ksd7w7jkx.1.

\bibitem{b42} N. Vallez, G. Bueno, O. Deniz, M. A. Rienda, and C. Pastor, "BUS-UCLM: Breast ultrasound lesion segmentation dataset," \textit{Mendeley Data}, version 1, 2024. doi: 10.17632/7fvgj4jsp7.1.

% \bibitem{b43} Wilfrido Gómez-Flores, Maria Julia Gregorio-Calasand Wagner Coelho de Albuquerque Pereira, "BUS-BRA: A Breast Ultrasound Dataset for Assessing Computer-aided Diagnosis Systems". \textit{Zenodo}, Mar. 13, 2023. doi: 10.5281/zenodo.8231412.

\bibitem{b44} M. H. Yap, G. Pons, J. Marti, S. Ganau, M. Sentis, R. Zwiggelaar, A. K. Davison, and R. Marti, "Automated breast ultrasound lesions detection using convolutional neural networks," \textit{IEEE Journal of Biomedical and Health Informatics}, 2017. doi: 10.1109/JBHI.2017.2731873.

\bibitem{b45} M. Alzubaidi, M. Agus, M. Makhlouf, F. Anver, K. Alyafeiand M. Househ, "Large-Scale Annotation Dataset for Fetal Head Biometry in Ultrasound Images", \textit{Data in Brief}, p. 109708, Mar. 2023, doi: 10.5281/zenodo.8265464.

\bibitem{b46} S. Leclerc, E. Smistad, J. Pedrosa, A. Ostvik, F. Cervenansky, F. Espinosa, T. Espeland, E. A. R. Berg, P.-M. Jodoin, T. Grenier, C. Lartizien, J. Dhooge, L. Lovstakken, and O. Bernard, "Deep learning for segmentation using an open large-scale dataset in 2D echocardiography," \textit{IEEE Transactions on Medical Imaging}, vol. 38, no. 9, pp. 2198–2210, Sept. 2019. doi: 10.1109/TMI.2019.2900516.

\bibitem{b47} J. Yang, X. Ding, Z. Zheng, X. Xu, and X. Li, "GraphEcho: Graph-Driven Unsupervised Domain Adaptation for Echocardiogram Video Segmentation," \textit{in Proceedings of the IEEE/CVF International Conference on Computer Vision (ICCV)}, 2023.

\bibitem{b48} A. Momot, "Common Carotid Artery Ultrasound Images," \textit{Mendeley Data}, version 1, 2022. doi: 10.17632/d4xt63mgjm.1.

\bibitem{b49} T. Wang, Z. Li, S. Bi, M. Huang, J. Zhang, J. Zhuang, Y. Shi, H. Fei, and X. Xu, "ImageCHD: A 3D computed tomography image dataset for classification of congenital heart disease," \textit{in Proceedings of the Medical Image Computing and Computer Assisted Interventions (MICCAI)}, Online, 2021.

\bibitem{b50} D. Ouyang, B. He, A. Ghorbani, N. Yuan, J. Ebinger, C. P. Langlotz, P. A. Heidenreich, R. A. Harrington, D. H. Liang, E. A. Ashley, and J. Y. Zou, "Video-based AI for beat-to-beat assessment of cardiac function," \textit{Nature}, vol. 580, no. 7802, pp. 252–256, Apr. 2020. doi: 10.1038/s41586-020-2145-8.

\bibitem{b51} C. D. Reddy, L. Lopez, D. Ouyang, J. Y. Zou, and B. He, "Video-Based Deep Learning for Automated Assessment of Left Ventricular Ejection Fraction in Pediatric Patients," \textit{Journal of the American Society of Echocardiography}, 2022. doi: 10.1016/j.echo.2023.01.015.

\bibitem{b52} "FAscicle Lower Leg Muscle Ultrasound Dataset." Available: https://kalisteo.cea.fr/index.php/fallmud/.

\bibitem{b53} K. S. Da Correggio, R. N. Galluzzo, L. O. Santos, F. S. M. Barroso, T. Z. L. Chaves, A. S. Onofre, and A. von Wangenheim, "Fetal Abdominal Structures Segmentation Dataset Using Ultrasonic Images," \textit{Mendeley Data}, version 1, 2023. doi: 10.17632/4gcpm9dsc3.1.

\bibitem{b54} V. Ashkani Chenarlogh, M. G. Oghli, A. Shabanzadeh, N. Sirjani, A. Akhavan, I. Shiri, H. Arabi, M. Sanei Taheri, and M. K. Tarzamni, "Fast and accurate U-Net model for fetal ultrasound image segmentation," \textit{Ultrasonic Imaging}, vol. 44, no. 1, pp. 25–38, Jan. 2022. doi: 10.1177/01617346211069882.

\bibitem{b55} B. Jieyun and O. ZhanHong, "Pubic Symphysis-Fetal Head Segmentation and Angle of Progression". \textit{Zenodo}, Apr. 20, 2023. doi: 10.5281/zenodo.7851339.

% \bibitem{b56} Q. He, S. Bano, J. Liu, et al., "Query2: Query over queries for improving gastrointestinal stromal tumour detection in an endoscopic ultrasound," \textit{Computers in Biology and Medicine}, vol. 152, p. 106424, 2023.

\bibitem{b57} Thomas L. A. van den Heuvel, Dagmar de Bruijn, Chris L. de Korteand Bram van Ginneken, "Automated measurement of fetal head circumference using 2D ultrasound images". \textit{Zenodo}, Jul. 27, 2018. doi: 10.5281/zenodo.1327317.

\bibitem{b58} R. Singla, C. Ringstrom, G. Hu, V. Lessoway, J. Reid, C. Nguan, and R. Rohling, "The open kidney ultrasound data set," in Proceedings of the International Workshop on Advances in Simplifying Medical Ultrasound, Cham, Switzerland: \textit{Springer Nature}, pp. 155–164, 2023.

\bibitem{b59} J. R. McLaughlan, L. Howell, and N. Ingram, "Lung ultrasound COVID phantom dataset used for training machine learning model," \textit{University of Leeds, Dataset}, 2024. doi: 10.5518/1485.

\bibitem{b60} W. Shao and W. Brisbane, "Micro-Ultrasound Prostate Segmentation Dataset". \textit{Zenodo}, Jan. 09, 2024. doi: 10.5281/zenodo.10475293.

% \bibitem{b61} Q. Zhao, S. Lyu, W. Bai, et al., "A multi-modality ovarian tumor ultrasound image dataset for unsupervised cross-domain semantic segmentation," \textit{arXiv preprint}, arXiv:2207.06799, 2022.

\bibitem{b62} J. Born, N. Wiedemann, M. Cossio, C. Buhre, G. Brändle, K. Leidermann, A. Aujayeb, M. Moor, B. Rieck, and K. Borgwardt, "Accelerating detection of lung pathologies with explainable ultrasound image analysis," \textit{Applied Sciences}, vol. 11, no. 2, p. 672, Jan. 2021. doi: 10.3390/app11020672.

\bibitem{b63} A. Kumar, K. Kotkar, K. Jiang, M. Bhimreddy, D. Davidar, C. Weber-Levine, S. Krishnan, M. J. Kerensky, R. Liang, K. K. Leadingham, D. Routkevitch, A. M. Hersh, K. Ashayeri, B. Tyler, I. Suk, J. Son, N. Theodore, N. Thakor, and A. Manbachi, "A novel open-source ultrasound dataset with deep learning benchmarks for spinal cord injury localization and anatomical segmentation," \textit{Scientific Reports}, vol. 15, art. no. 33192, 2025. doi: 10.1038/s41598-025-16275-z.

\bibitem{b64} Zachary M. C. Baum, Shaheer U. Saeed, Zhe Min, Yipeng Huand Dean C. Barratt, "MR to Ultrasound Registration for Prostate Challenge - Dataset". \textit{Zenodo}, Jun. 05, 2023. doi: 10.5281/zenodo.8004388.

\bibitem{b65} Y. Guo, X. Duan, C. Wang, and H. Guo, "Segmentation and recognition of breast ultrasound images based on an expanded U-Net," \textit{PLOS ONE}, vol. 16, no. 6, e0253202, 2021. doi: 10.1371/journal.pone.0253202.

\bibitem{b66} M. Krönke, C. Eilers, D. Dimova, M. Köhler, G. Buschner, L. Schweiger, L. Konstantinidou, M. Makowski, J. Nagarajah, N. Navab, W. Weber, and T. Wendler, "Tracked 3D ultrasound and deep neural network-based thyroid segmentation reduce interobserver variability in thyroid volumetry," \textit{PLOS ONE}, July 29, 2022. doi: 10.1371/journal.pone.0268550.

\bibitem{b67} F. Marzola, N. van Alfen, J. Doorduin, and K. Meiburger, "Dataset for ‘Deep learning segmentation of transverse musculoskeletal ultrasound images for neuromuscular disease assessment’," \textit{Mendeley Data}, version 1, 2021. doi: 10.17632/3jykz7wz8d.1.

\bibitem{b68} X. B. H. L. K, STU-Hospital, GitHub repository, followed commit version, year. Available: https://github.com/xbhlk/STU-Hospital.

\bibitem{b69} Stanford AIMI Shared Datasets, "Dataset ID a72f2b02-7b53-4c5d-963c-d7253220bfd5," Stanford AIMI Shared Datasets. Available: https://stanfordaimi.azurewebsites.net/datasets/a72f2b02-7b53-4c5d-963c-d7253220bfd5.

% \bibitem{b70} A. Montoya, Hasnin, kaggle446, Shirzad, W. Cukierski, and yffud, "Ultrasound Nerve Segmentation," \textit{Kaggle}, 2016. Available: https://www.kaggle.com/competitions/ultrasound-nerve-segmentation.

\bibitem{b71} Orvile, "Ultrasound Fetus Dataset," \textit{Kaggle}. Available: https://www.kaggle.com/datasets/orvile/ultrasound-fetus-dataset.


\bibitem{b72} M. Bertalmio, A. L. Bertozzi, and G. Sapiro, "Navier–Stokes, fluid dynamics, and image and video inpainting," \textit{in Proceedings of the 2001 IEEE Computer Society Conference on Computer Vision and Pattern Recognition (CVPR 2001)}, Kauai, HI, USA, 2001, pp. I–I. doi: 10.1109/CVPR.2001.990497.

\bibitem{b73} K. He, X. Chen, S. Xie, Y. Li, P. Dollár, and R. Girshick, "Masked autoencoders are scalable vision learners," \textit{in Proceedings of the IEEE/CVF Conference on Computer Vision and Pattern Recognition (CVPR)}, New Orleans, LA, USA, pp. 16000–16009, 2022.

\bibitem{b74} Catalin Stoean, Ruxandra Stoean, Mircea Hotoleanu, Dominic Gabriel Iliescu, Ciprian Patru, Rodica Nagy, "An assessment of the usefulness of image pre-processing for the classification of first trimester fetal heart ultrasound using convolutional neural networks", \textit{25th International Conference on System Theory, Control and Computing (ICSTCC 2021)}, October 20-23, IEEE, pp. 242-248, 2021.

\bibitem{b75} Ruxandra Stoean, Dominic Iliescu, Catalin Stoean, Vlad Ilie, Ciprian Patru, Mircea Hotoleanu, Rodica Nagy, Dan Ruican, Rares Trocan, Andreea Marcu, Miguel Atencia and Gonzalo Joya, "Deep Learning for the Detection of Frames of Interest in Fetal Heart Assessment from First Trimester Ultrasound". In: Rojas I., Joya G., Català A. (eds) \textit{Advances in Computational Intelligence}. IWANN 2021. Lecture Notes in Computer Science, vol 12861. Springer, Cham, 2021. 

\bibitem{b76} A. Kirillov, E. Mintun, N. Ravi, H. Mao, C. Rolland, L. Gustafson, T. Xiao, S. Whitehead, A. C. Berg, W.-Y. Lo, P. Dollár, and R. Girshick, "Segment anything," \textit{in Proc. IEEE/CVF Int. Conf. Comput. Vis. (ICCV)}, 2023, pp. 4015–4026.

\bibitem{b77} C. G. Predoi, C. Grigoriu, R. Vladescu, and A. E. Mihart, "Placental damages in preeclampsia – from ultrasound images to histopathological findings," \textit{J. Med. Life}, vol. 8, Spec. Issue, pp. 62–65, 2015.

\bibitem{b78} M. C. Walker, I. Willner, O. X. Miguel, M. S. Q. Murphy, D. El-Chaâr, F. Moretti, A. L. J. Dingwall Harvey, R. Rennicks White, K. A. Muldoon, A. M. Carrington, S. Hawken, and R. I. Aviv, "Using deep-learning in fetal ultrasound analysis for diagnosis of cystic hygroma in the first trimester," \textit{PLoS One}, vol. 17, no. 6, e0269323, 2022, doi: 10.1371/journal.pone.0269323.

\bibitem{b79} B. Yakıştıran, O. Altınboğa, E. Canpolat, E. Ş. Çakar, Ş. Çelen, A. T. Çağlar, and Y. Engin Üstün, "Analysis of cystic hygroma diagnosed in the first trimester: Single-center experience," \textit{J. Turkish German Gynecol. Assoc.}, vol. 21, no. 2, pp. 107–110, 2020, doi: 10.4274/jtgga.galenos.2019.2019.0032.

\bibitem{b80} S. Chetty, "Multicystic dysplastic kidney," \textit{Am. J. Obstet. Gynecol.}, vol. 225, no. 5, pp. B21–B22, 2021, doi: 10.1016/j.ajog.2021.06.046.

\bibitem{b81} O. X. Miguel, E. Kaczmarek, I. Lee, R. Ducharme, A. L. J. Dingwall-Harvey, R. R. White, B. Bonin, R. I. Aviv, S. Hawken, C. M. Armour, K. Dick, and M. C. Walker, "Deep learning prediction of renal anomalies for prenatal ultrasound diagnosis," \textit{Sci. Rep.}, vol. 14, p. 9013, 2024, doi: 10.1038/s41598-024-59248-4.

\bibitem{b82} J. B. Russ, S. Agarwal, C. Venkatesan, B. Scelsa, B. Vollmer, T. Tarui, A. C. Pardo, M. E. Lemmon, S. B. Mulkey, A. R. Hart, U. D. Nagaraj, J. A. Kuller, M. T. Whitehead, J. L. Cohen, J. S. Gebb, O. A. Glenn, M. E. Norton, and D. Gano, "Fetal malformations of cortical development: review and clinical guidance," \textit{Brain}, vol. 148, no. 6, pp. 1888–1903, 2025, doi: 10.1093/brain/awaf094.


\end{thebibliography}
\end{document}